\documentclass{aa}
\usepackage{natbib,twoopt}
\usepackage{xcolor}
\usepackage{subcaption}
\usepackage[breaklinks=true, colorlinks=true, linkcolor=blue, citecolor=blue]{hyperref} %% to avoid \citeads line fills
\bibpunct{(}{)}{;}{a}{}{,}             %% natbib format for A&A and ApJ
\makeatletter
\newcommandtwoopt{\citeads}[3][][]{\href{http://adsabs.harvard.edu/abs/#3}%
    {\def\hyper@linkstart##1##2{}%
     \let\hyper@linkend\@empty\citealp[#1][#2]{#3}}}
  \newcommandtwoopt{\citepads}[3][][]{\href{http://adsabs.harvard.edu/abs/#3}%
    {\def\hyper@linkstart##1##2{}%
     \let\hyper@linkend\@empty\citep[#1][#2]{#3}}}
  \newcommandtwoopt{\citetads}[3][][]{\href{http://adsabs.harvard.edu/abs/#3}%
    {\def\hyper@linkstart##1##2{}%
     \let\hyper@linkend\@empty\citet[#1][#2]{#3}}}
  \newcommandtwoopt{\citeyearads}[3][][]%
    {\href{http://adsabs.harvard.edu/abs/#3}
    {\def\hyper@linkstart##1##2{}%
     \let\hyper@linkend\@empty\citeyear[#1][#2]{#3}}}
\makeatother
\usepackage{graphicx}
\usepackage{multirow}
\usepackage{siunitx}
\usepackage{txfonts}

\usepackage{graphicx}   % Including figure files
\usepackage{amsmath}    % Advanced maths commands
\usepackage{amssymb}    % Extra maths symbols
\usepackage[normalem]{ulem}
\usepackage{placeins}

\usepackage{color}

\newcommand{\rvir}{R_{200}}
\newcommand{\rfive}{R_{500}}

\newcommand{\cMpc}{h^{-1}\,{\rm cMpc}}

\newcommand{\Mpc}{{\rm Mpc}}
\newcommand{\kpc}{{\rm kpc}}

\newcommand{\msun}{{\rm M}_{\odot}}
\newcommand{\msunh}{h^{-1}\,{\rm M}_{\odot}}

\newcommand{\mvir}{M_{\rm tot,200}}
\newcommand{\mfive}{M_{\rm tot,500}}
\newcommand{\keV}{{\rm keV}}

\newcommand{\Magneticum}{\textsc{Magneticum}}

\newcommand{\Subfind}{\textsc{SubFind}}

\begin{document}

\title{The full iron budget in simulated galaxy clusters: \\
the chemistry between gas and stars}

\author{Veronica~Biffi\inst{1,2}\thanks{e-mail:         
    \href{mailto:veronica.biffi@inaf.it}{\tt veronica.biffi@inaf.it}}
    \and Elena~Rasia\inst{1,2,3}
    \and Stefano~Borgani\inst{1,2,4,5,6}
    \and Simona~Ghizzardi\inst{7}
    \and Umberto~Maio\inst{1,2}
    \and Klaus~Dolag\inst{8,9}
    \and Fabio~Gastaldello\inst{7}
    \and Luca~Tornatore\inst{1,2}
    }
    
    \titlerunning{Galaxy cluster iron budget}
    \authorrunning{V. Biffi et al.}

\institute{INAF --- Osservatorio Astronomico di Trieste, via Tiepolo 11, I-34143 Trieste, Italy
    \and
    IFPU --- Institute for Fundamental Physics of the Universe, Via Beirut 2, I-34014 Trieste, Italy
    \and
    Department of Physics; University of Michigan, 450~Church~St, Ann Arbor, MI~48109, USA
    \and
    Department of Physics, University of Trieste, via G. Tiepolo 11, I-34131 Trieste, Italy
    \and
    ICSC --- Italian Research Center on High Performance Computing, Big Data and Quantum Computing, via Magnanelli 2, 40033, Casalecchio di Reno, Italy
    \and
    INFN --- Istituto Nazionale di Fisica Nucleare, via Valerio 2, I-34127, Trieste, Italy 
    \and
    INAF --- Istituto di Astrofisica Spaziale e Fisica Cosmica di Milano, via A. Corti 12, I-20133 Milano, Italy
    \and
    Universit\"ats-Sternwarte, Fakult\"ut f\"ur Physik, Ludwig-Maximilians-Universit\"at M\"unchen, Scheinerstr.1, 81679 M\"unchen, Germany 
    \and
    Max-Planck-Institut f\"ur Astrophysik, Karl-Schwarzschild-Straße 1, 85741 Garching, Germany
    }
\date{Received; accepted}

\abstract
    %\abstract{}{}{}{}{} % 5 {} token are mandatory
    % context heading (optional)
    % {} leave it empty if necessary  
    {
    The presence of heavy chemical elements such as iron in the intra-cluster medium (ICM) of galaxy clusters is a signpost of the interaction between the gas and stellar components. Observations of the ICM metallicity in present-day massive systems pose however a challenge to the underlying assumption that the cluster galaxies have produced the amount of iron enriching the ICM.
    }
  % aims heading (mandatory)
    {
    The goal of this study is to evaluate the iron share between ICM and stars within simulated galaxy clusters 
    with the twofold aim of investigating the origin of possible differences with respect to observational findings, and of shedding light on the observed excess of iron on the ICM with respect to expectations based on the observed stellar population.
    }
  % methods heading (mandatory)
    {
    We evaluate the iron mass in gas and stars in a sample of $448$ simulated systems with masses $\mfive > 10^{14}\msun$ at $z=0.07$. These are extracted from the high-resolution $(352 \,\cMpc)^3$ volume of the \Magneticum{} cosmological hydrodynamical simulations. 
    We compare our results with observational data of low-redshift galaxy clusters.
    }
  % results heading (mandatory)
    {
    The iron share in simulated clusters features a shallow dependency on the total mass and its value is on average close to unity. In the simulated most massive systems, the iron share is thus smaller than observational values by almost an order of magnitude.
    The dominant contribution to this difference is related to the stellar component, whereas the ICM chemical properties show an overall good agreement with observations. 
    We find larger stellar mass fractions in simulated massive clusters, which in turn yield larger stellar iron masses, compared to observational data.
    }
  % conclusions heading (optional), leave it empty if necessary 
    {
    Consistently with the modelling, this study confirms that the stellar content within simulated present-day massive systems is responsible for the ICM metal enrichment.
    Alleviating the stellar-mass discrepancy between simulations and observations will be crucial for a definitive assessment of the iron budget in galaxy clusters.
    }

\maketitle

\section{Introduction}

Chemical elements heavier than helium, often referred to as metals, constitute a tiny fraction of the total mass budget in galaxy clusters. Yet, their presence in the hot intra-cluster medium (ICM) is the signpost of the interplay between stars and gas from galaxy to cluster scales.
In the self-similar scenario of structure formation~\cite[][]{gg1972,white1978,peebles1980}, ICM thermodynamical properties, such as temperature, density and pressure, can be connected directly to the system gravitating mass, at first order~\cite[][]{kaiser1986}. 
Deviations from self-similarity are typically related to the impact of additional physical processes, such as star formation and energetic feedback, released by either stellar sources or super-massive black holes powering active galactic nuclei (AGN).
Un-related to the self-similar scenario, ICM chemical features are essentially connected to the presence of a stellar component that synthesized the metals and enriched the gas.
Nonetheless, observational data and simulation results of the last decades all point towards a consistent picture in which the metallicity profiles of the hot gas in galaxy groups and clusters are remarkably self-similar and global chemical properties do not vary much across different systems~\cite[see for instance reviews by][and references therein]{mernier2018Rv,biffi2018Rv,gastaldello2021} 
and in time~\cite[$z\lesssim2$; e.g.][]{baldi2012}.

In this framework, the gas chemical enrichment from groups to clusters of galaxies features on average a spatially homogeneous distribution of the metals at large cluster-centric distances, marked by the flatness of the metallicity radial profiles beyond the central regions~\cite[][]{degrandi2001,leccardi2008,werner2013,mernier2017,urban2017,ghizzardi2021}. 
In the inner regions, spatially resolved X-ray observations have revealed in some cases asymmetries in the 2D metallicity maps of single clusters, thus indicating the presence of azimuthal variation. Dedicated studies have shown that this is in most of the cases related to metal-rich outflows associated to AGN-powered radio jets or to sloshing and mergers~(\citealt{rebusco2005,david2008,osullivan2011,ghizzardi2014,kirkpatrick2011,kirkpatrick2015}; see also numerical studies by~\citealt{gaspari2011}).
Instead in the outermost regions, the radial metallicity profiles show spatial homogeneity~\cite[][]{werner2013,simionescu2015,vogelsberger2018}, and this result is not only valid for iron (Fe), but also for other chemical elements, produced by different enrichment sources like  supernovae (SN) Type Ia and core-collapse (SNIa and SNcc, respectively) or intermediate- and low-mass stars during their asymptotic giant branch (AGB) phase~\cite[][]{matsushita2013,simionescu2015,ezer2017}.
The homogeneity of gas enrichment is also preserved across different clusters. This is marked by a relative low scatter on the ICM metallicity profiles in the outskirts~\cite[][]{urban2017,mernier2017,gastaldello2021,ghizzardi2021,sarkar2022}, and 
by a shallow dependency of global metallicity on the system temperature (i.e.\ total mass)
over a large mass range~\cite[][]{andreon2012b,truong2019,mernier2018a,mernier2018b}.
This picture is furthermore valid in clusters of the local Universe and up to redshift $z\sim 2$, thus indicating little evolution over time especially beyond the core region~\cite[][]{baldi2012,ettori2015,mcdonald2016,mantz2017,liu2020,flores2021}.
Most simulation results agree with this global picture~\cite[][]{biffi2018Rv}, although differences persist in some aspects due to the impact of the detailed galaxy formation modelling onto the ICM chemical pattern~\cite[see for instance recent results by][]{hough2024}. 
In particular, the shallow variation of the gas metallicity across the mass range from groups to clusters~\cite[][]{truong2019,pearce2021,nelson2024}
is still debated. Recent findings by~\cite{braspenning2024} and~\cite{padawer-blatt2025}
rather indicate a stronger mass dependence of the radial metallicity profiles --- although the difference between groups and clusters remains within a factor of $\sim2$.

The aforementioned evidences about the uniformity and time-invariance of the ICM enrichment point towards an early enrichment ($z\gtrsim 3$) of the gas by the stellar populations of galaxies~\cite[][]{renzini2004}, followed by an efficient and broad displacement of metal-rich gas due to the effect of an efficient feedback mechanism at early times --- for instance from AGNs~\cite[][]{werner2013}, as also supported by simulation studies (e.g.~\citealt{fabjan2010,biffi2017,biffi2018}; see also~\citealt{mccarthy2011}, on the impact of early AGN feedback on simulated cluster X-ray properties).

The connection to the stellar assembly process and total stellar content in groups and clusters of galaxies is thus key to fully explain the metal cycle in cosmic structures~\cite[][]{renzini1993}.
A complete census of the metals in massive systems is nonetheless very challenging to be achieved from data. 
The direct measurement of the iron mass in cluster gas and stars requires both ICM abundances and stellar masses to be ideally measured out to large radii, so as to comprise the whole cluster baryonic content. A representative region to consider is the one encompassed by $\rvir$ (or $\rfive$), within which the average matter density of the system is $200$ (or $500$) times the critical density of the Universe at the system's redshift.
For what concerns the gas component, observational measurements of the ICM metallicity suggest that potential systematic issues can affect the abundance measurements in the cluster outskirts~\cite[][]{molendi2016}.
The main limitation remains in fact to cover a large fraction of the cluster volume for a representative sample of systems. In the last decade, the intermediate and outer regions of galaxy clusters have been probed for an increasing number of objects, for which the iron abundance profiles have been derived from X-ray Suzaku~\cite[e.g.][]{fujita2008,urban2011,simionescu2011,urban2017} and XMM-Newton data (see e.g.~\citealt{ghizzardi2021} and the review by~\citealt{mernier2018Rv}).
These progresses have been crucial to refine ICM iron mass estimates, obtained by integrating the abundance profile~\cite[e.g.][]{degrandi2004}, instead of assuming a constant average ICM abundance and relying on the correlation between gas iron mass and total gas mass~\cite[e.g.][]{renzini1993}.

Regarding the stellar component, the common practise is to calculate the iron mass from the total stellar mass assuming an average solar abundance~\cite[][]{renziniandreon2014,maoz2017,ghizzardi2021}.
The main source of uncertainty thus resides in the stellar mass estimate, typically derived from optical/near-IR observations of the brightest cluster galaxy (BCG) and satellite galaxies in the cluster. This measurement is complicated by the low surface brightness of extended light beyond the central regions occupied by BCG and by the possible presence of multiple stellar populations~\cite[][]{mitchell2013}.
Also, the uncertain contribution of the intra-cluster light (ICL) to the total stellar light of the cluster adds a non-negligible source of uncertainty to the stellar mass determination~\cite[][]{mihos2019} and causes substantial differences among the various observational analyses. Typically, values range from $\sim 10\%$ to $\sim 50\%$~\cite[][]{zibetti2005, gonzalez2007, montes2018, deoliveira2022}, consistent with recent results from Euclid observations of the Perseus cluster indicating a BCG+ICL fraction of $\sim 35\%$~\cite[][]{kluge2024}. 
Another source of uncertainty is related to the assumptions needed to infer stellar masses from broad-band photometry, especially the assumed stellar initial mass function (IMF),
the measurement of the total galaxy luminosities \citep[e.g., see discussion in][]{kravtsov2018},
as well as different apertures and forward modellings~\citep{price2017, deGraaff2022}.

Given the observational estimates of the iron budget in clusters, 
attempts to reconcile the observed ICM metallicity and the expected enrichment from the observed stellar content have failed so far~\cite[as already discussed in early studies by e.g.\ ][]{vigroux1977}. 
The amount of iron observed in the gas phase is consistent with expectations based on the observed stellar population in the group regime.
Differently, at the scale of massive clusters, the ICM iron budget inferred from the observed abundance profiles seems to exceed significantly the amount that the stars in cluster galaxies could have reasonably produced, based on standard nucleosynthesis and stellar evolution models~\cite[e.g.][]{arnaud1992,loewenstein2006,loewestein2013,renziniandreon2014,ghizzardi2021}.
This represents the so-called iron conundrum~\cite[as formulated by][]{renziniandreon2014} and essentially follows from three observational evidences, namely (i) the increase of gas fraction with total mass; (ii) the relatively flat metallicity-mass relation; (iii) the decrease in stellar fraction with system mass.

In massive objects, in fact, observations find a significant drop in stellar mass fraction which is not however accompanied by a drop in ICM metallicity, thus generating a significant tension with the stellar nucleosynthesis predictions. 
In these clusters, the ratio between the iron mass in the ICM and the iron mass locked into stars, that is the iron share, is several times larger than expected. Estimates of the effective efficiency with which stars produce iron in clusters have also been performed via the iron yield calculation --- namely the ratio between the total iron mass in baryons (ICM and stars) and the mass of gas that turned into the stellar component --- still finding a discrepancy with the theoretical predictions from stellar evolution models~\cite[][]{renziniandreon2014,ghizzardi2021}. 
Several solutions have been proposed in the last decades to reconcile this puzzling observational findings with theoretical expectations, among which an incorrect assumption on the stellar IMF in clusters, higher SNIa rates at high redshifts or within cluster environment, a different efficiency of metal production and release into the ICM, or even additional stellar populations contributing to the ICM enrichment such as Population III stars and pair-instability SN~\cite[e.g.][]{loewenstein2001,loewestein2013,bregman2010,morsony2014,maoz2017,blackwell2022}.
Recently, \cite{molendi2024} proposed an observation-based model for cluster chemical enrichment that rather suggests the Fe conundrum to arise from the underestimation of observed stellar masses and to the different distributions of gas and stars in massive systems.

In this paper, we investigate the relation between stellar content and ICM enrichment in simulated clusters
extracted from the \Magneticum{} state-of-the-art cosmological hydrodynamical simulations, which reproduce fairly well the thermodynamical and chemical properties of the observed populations of galaxy groups and clusters.
A description of the simulation run and of the cluster sample employed is provided in Sec.~\ref{sec:sims}.
In particular, we evaluate directly the iron budget in the ICM and stars in the simulated systems to assess whether there is any conundrum in simulations. 
Sec.~\ref{sec:iron_share} illustrates our results on the iron share in comparison to observations~\cite[e.g.\ by][]{renziniandreon2014,ghizzardi2021}.
In Sec.~\ref{sec:fe_mass}, we further inspect the total iron content of ICM and stellar components separately (Sections~\ref{sec:iron_budget_ICM} and~\ref{sec:iron_budget_stars}, respectively).
In order to investigate the origin of the difference between simulations and observations in terms of gas and stellar component separately, we primarily focus on the results by~\cite{ghizzardi2021}, based on the XMM Cluster Outskirts Project~\cite[X-COP;][]{eckert2017} dataset. This is in fact a recent representative sample of massive systems (with total mass enclosed within $\rfive$, $\mfive > 3 \times 10^{14}\msun$), for which ICM iron abundance profiles are measured out to $\rfive$, and corrected for a systematic error that has affected most of previous Fe abundance measurements in cluster outskirts (introduced by the Fe L-shell emission that can bias the fit and the resulting abundance measurements in those outer low-surface brightness regions). For most of the X-COP systems stellar mass profiles are also available from optical data~\cite[][]{vanderburg2015}, which allowed~\cite{ghizzardi2021} to derive a detailed estimate of the iron budget in ICM and galaxies within $\rfive$.
The connection between ICM enrichment level and star formation efficiency (traced by the stellar-to-gas mass fraction) is further investigated in Sec.~\ref{sec:SF_Z}. 
In Sec.~\ref{sec:iron_yield} we show results on the efficiency with which stars produce iron in simulated clusters, and compare them to observational estimates.
Our results and related limitations are discussed in Sec.~\ref{sec:discussion}. We summarise our main results and conclusions in Sec.~\ref{sec:conclusion}.

\section{Simulations}\label{sec:sims}

In the following we present our results based on the study of the most massive haloes in a large-volume cosmological box from the \Magneticum{} \textsc{Pathfinder\footnote{\texttt{www.magneticum.org}}} simulation set (hereafter, \Magneticum).
The \Magneticum{} simulations are performed with an improved version of \textsc{Gadget~3}, a non-public version of the Tree-PM/Smoothed-Particle-Hydrodynamics (SPH) code \textsc{Gadget~2}~\cite[][]{springel2005}.
The code includes an updated SPH scheme, based on the inclusion of a time-dependent artificial viscosity and of artificial thermal diffusion~\cite[][]{dolag2004,dolag2005,beck2016}, as well as higher-order interpolation kernels~\cite[][]{dehnen2012} and passive magnetic fields~\cite[][]{dolagMHD2009}.
A variety of physical processes driving the evolution of baryons is also modelled. 
Gas thermal properties are affected by the heating from a uniform time-dependent ultraviolet (UV) background~\cite[][]{haardt2001} and the radiative losses due to metallicity-dependent cooling processes~\cite[][]{wiersma2009}.
Gas parcels denser than $\sim 0.1\,\rm cm^{-3}$ become eligible to form stars according to the sub-resolution model for star formation by~\cite{springel2003}. According to this model star-forming particles having density exceeding the above threshold value are assumed to describe a multi-phase interstellar medium, in which a cold and a hot phase coexist in pressure equilibrium, with the former representing the reservoir for star formation.
Energy feedback is modelled both from stellar sources, in the form of thermal and kinetic feedback (with a fixed galactic-wind velocity of $v_{w} = 350\,{\rm km\,s}^{-1}$)~\cite[][]{springel2003,tornatore2010}, and from AGNs, as thermal feedback powered by gas accretion onto super-massive black holes (SMBHs).
The evolution of SMBHs is treated within the simulations according to the models by~\cite{springeldimatteo2005}  and~\cite{dimatteo2005}, with further modifications by~\cite{fabjan2010}.
Stellar evolution and chemical enrichment are included according to the model implemented by \cite{tornatore2007}. Specifically, metals are produced by SNIa,  
SNcc, 
and AGB stars, for which the \Magneticum{} simulations assume stellar yields by 
\cite{thielemann2003},
\cite{WW1995}, 
and~\cite{vandehoek1997}, 
respectively.
These three enrichment channels are associated to stellar particles that represent simple stellar populations with a IMF according to~\cite{chabrier2003} and mass-dependent lifetimes by~\cite{padovani1993}.
Eleven different chemical elements, i.e.\ H, He, C, Ca, O, N, Ne, Mg, S, Si, and Fe, are directly traced by the simulation code.
In particular, iron is mostly produced by long-lived stars and, in our simulations, we implement the single-degenerate scenario to account for metal production by SNIa events \cite[see e.g.][]{matteucci2001, greggio2005}.

The cosmology adopted in \Magneticum{} is consistent with the $7^{\rm th}$-year data release by the Wilkinson Microwave Anisotropy Probe~\cite[][]{wmap7}, with the
Hubble parameter at present time set to $H_0=100 \times h$[{km\ s$^{-1}$ Mpc$^{-1}$}] with $h=0.704$, 
the normalisation of the fluctuation amplitude at
$8\,h^{-1}\Mpc$ equal to $\sigma_8 = 0.809$
and the density parameters equal to $\Omega_M=0.272$, $\Omega_\Lambda=0.728$ and $\Omega_b=0.0451$, for matter,
dark energy and baryons, respectively.

A plethora of studies based on the \Magneticum{} simulations reported a number of consistent properties of the simulated clusters and galaxies compared to observational findings~\cite[see][]{dolag2025}.
In the galaxy regime, global and X-ray properties of normal galaxy and AGN populations have been successfully compared to observations~\cite[see for instance,][]{hirschmann2014,schulze2018,biffi2018,SVZ2023,rhitarsic2024}.
In particular, studies by \cite{teklu2017} and \cite{remus2017} have shown how the stellar mass function, the relation between stellar and total mass and the properties of central galaxies are in overall agreement with observations, especially in the intermediate mass range (halo masses of $\mvir \sim 10^{12}$--$10^{13}\msun$).
At cluster scales, several studies showed that \Magneticum{} haloes successfully match observed gas properties, such as scaling relations, pressure profiles, and X-ray properties~\cite[e.g.][]{biffi2013, gupta2017, ragagnin2019,biffi2022,zuhone2023,marini2024,bahar2024}.
More importantly for this work, the \Magneticum{} simulations yielded results on the chemical enrichment of the gas from galaxy to cluster scales that are in good agreement with observations, as reported by~\cite{dolag2017}, \cite{biffi2018Rv} and~\cite{kudritzki2021}.

\subsection{Dataset}

Among the various cosmological volumes and resolutions available in the \Magneticum{} suite, we focus on the {\it Box2} run carried out at high resolution ({\it hr}), which encompasses a cubic periodic volume of size $(352\,\cMpc)^3$ resolved with $2\times 1584^3$ particles. 
This corresponds to $m_{\rm DM}=6.9\times 10^8\,\msunh$ and $m_{\rm gas}=1.4\times 10^8\,\msunh$, for the mass of dark matter and gas particles, respectively.

Haloes in the \Magneticum{} simulation are identified using the \Subfind~algorithm \citep{springel2001, dolag2009}.
The centre of haloes is positioned on the most-bound particle location and characteristic radii and masses are computed for each halo with the spherical over-density method. 
We define $R_{\Delta}$ as the radius enclosing the sphere in which the average matter density is $\Delta$ times the critical density of the Universe at the redshift of the system, and the total mass enclosed within that radius is $M_{\rm tot,\Delta}\equiv M_{\rm tot}(<R_{\Delta})$.
In the following, we will mainly use $\rfive$ and $\rvir$, and the masses therein, namely $\mfive$ and $\mvir$.
By using $M_{\rm tot,\Delta}$ or referring to the ``total mass'', we always refer to $M_{\rm tot,\Delta}$, namely the true total mass of the system within the radius $R_\Delta$, as defined above.
Similarly, we define the total gas and stellar masses ($M_{\rm gas,\Delta}$ and $M_{\rm *,\Delta}$, respectively), computed by summing up all the masses of the gas and stellar particles located within $R_\Delta$, with no distinction between main halo and substructures (unless otherwise specified, see Sec.~\ref{sec:mstar_estimate}).
Gravitationally-bound substructures within each main halo can be nonetheless identified by the \Subfind~algorithm, thus 
allowing to distinguish between central and satellite galaxies.

In order to study the iron budget in gas and stars in massive systems, we focus on a simulation snapshot of {\it Box2/hr} corresponding to $z=0.07$. We extract therefrom a sample of 448 objects, comprising all the systems with $\mfive > 10^{14}M_\odot$.

\subsection{Methods}
For the analysis of the gas chemical properties that would be derived from X-ray observables, namely iron gas masses and metallicities, we restrict to the hot and tenuous gas phase, which is expected to emit at X--ray energies.
Specifically, we select all gas particles that are not star-forming, and the multi-phase star forming particles whose cold fraction is smaller than $10\%$. Selected gas particles must have a temperature in the range $ 0.3 < T [\keV] < 50 $. 
The lower temperature limit is used to restrict to the hot X-ray-emitting phase of the ICM. Changing this threshold has little effect in the mass regime explored, especially considering mass-weighted averages.
The upper limit aims at excluding spurious ultra-energetic particles that are occasionally heated to such high temperatures by the AGN feedback, still need to thermalize in the ICM by shock heating, and would then provide an unrealistic contribution to X-ray emission.
The X-ray gas component selected this way constitutes more than $95\%$ of the total gas mass (median value in the sample), within both $\rfive$ and $\rvir$.
To distinguish the two cases, we use the label ``ICM'' when referring to this X-ray hot phase (e.g.\ $M_{\rm Fe, ICM}$), and ``gas'' when all gaseous components are considered instead (e.g.\ $M_{\rm gas,\Delta}$ or $f_{\rm gas}$).
For what concerns the total gas mass, we verified that using all the gas particles or only the X-ray component has no appreciable impact on the relations discussed below.

The total iron content is computed for gas and stellar components by summing up the iron mass contained in X-ray gas and stellar particles located within the considered spherical region, that is $M_{\rm Fe,ICM}(<R_\Delta)$ and $M_{\rm Fe,*}(<R_\Delta)$, respectively.
For the stellar component, an observational-like estimate is also provided in Sec.~\ref{sec:iron_share}, Eq.~\eqref{eq:iron_mass_obs}, for the purpose of comparing with observational results.

As for the reference solar iron abundance, we adopt the value by~\cite{asplund2009} throughout the paper, that is $Z_{\rm Fe,\odot}=3.16 \times 10^{-5}$ (number fraction relative to hydrogen).

\section{Results} \label{sec:results}
In the following we present our main findings about the iron share (Sect.~\ref{sec:iron_share}) and the iron budget in the gaseous and stellar components (Sect.~\ref{sec:fe_mass}).
This analysis will allow us to connect star formation properties with the chemical enrichment of the ICM (Sect.~\ref{sec:SF_Z}). 
In  Sec.~\ref{sec:iron_yield} we additionally inspect the efficiency with which stars produce iron in our simulated clusters by investigating the effective iron yield.

\subsection{Iron share} \label{sec:iron_share}

\begin{figure}
\centering
\includegraphics[width=.99\columnwidth,trim=10 0 10 10,clip]{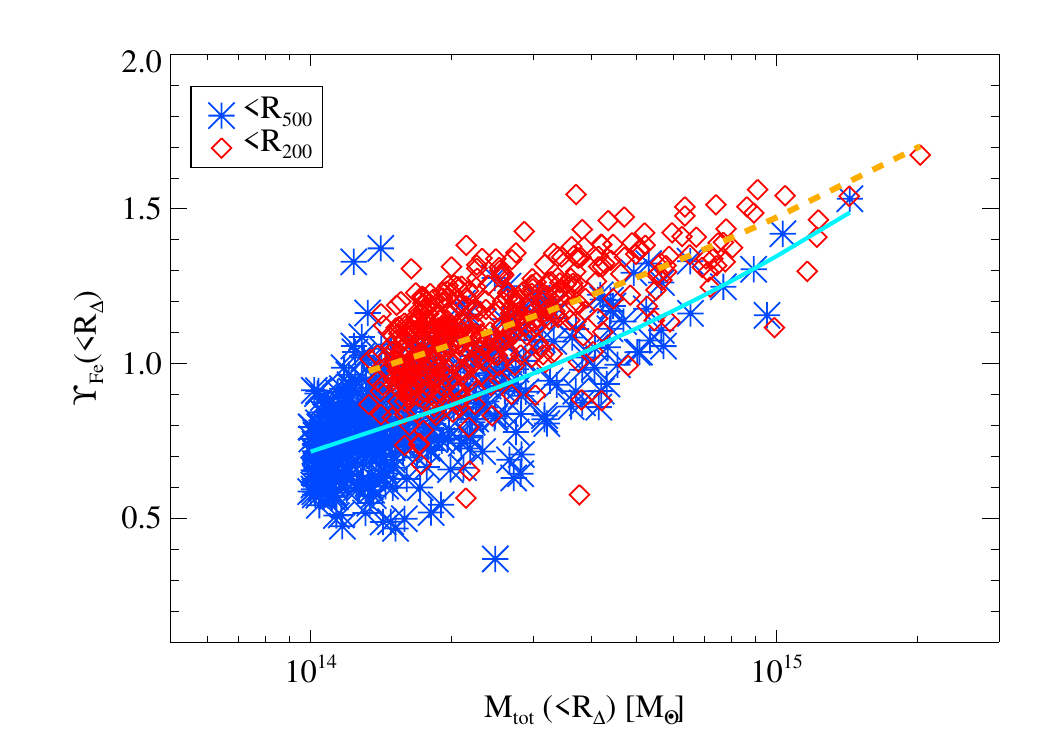}     
    \caption{Iron share between ICM and stellar component as a function of the system total mass. On both axes, quantities are computed either within $\rfive$ (blue asterisks) or within $\rvir$ (red diamonds). 
    The two lines correspond to the linear best fit of the two datasets in the log-log space.
    \label{fig:Fe-share}}
\end{figure}

The iron share~\cite[][]{renziniandreon2014,ghizzardi2021}, $ \Upsilon_{\rm Fe}$, is defined as the ratio between the mass of iron contained in the ICM ($M_{\rm Fe,ICM}$) and the mass of iron locked into stars ($M_{\rm Fe,*}$) in the same cluster volume, for instance, the spherical region enclosed within a distance $ R $ from the centre:
\begin{equation}
    \Upsilon_{\rm Fe} (<R) = \frac{M_{\rm Fe,ICM}(<R)} {M_{\rm Fe,*} (<R)}.
    \label{eq:iron_share}
\end{equation}
This quantity depends on the effective iron yields of the stars that enriched the ICM during the cluster assembly, which is, in turn, sensitive to assumptions about the IMF, stellar yields and, possibly, dynamical and feedback history of the system.
In cluster simulations, the iron mass within a given radius can be directly measured  and compared to available observational findings.
In Fig.~\ref{fig:Fe-share}, results for our simulated haloes are displayed as a function of the cluster total mass, 
for $\rfive$ and $\rvir$ (as in the legend).
In both cases, $ \Upsilon_{\rm Fe} $ increases with mass in the whole range between $ 10^{14} $ and $ 2 \times 10^{15}\, \rm M_\odot $.
For a more quantitative comparison of the iron-share to total mass relation, 
we fit the two datasets in Fig.~\ref{fig:Fe-share} with a linear relation in the log-log space of the form $\log(Y_\Delta) = A_\Delta + B_\Delta \times \log(X_\Delta)$, with $Y_\Delta=\Upsilon_{\rm Fe} (<R_\Delta)$ and $X_\Delta = M_{\rm tot,\Delta}/M_\odot$, for  $\Delta=500$ or $\Delta=200$. 
From the best-fit results, we find that the two distributions have relatively similar
slopes 
($B_{500}=0.28\pm 0.02$ and $B_{200}=0.20\pm 0.01$),
and normalisations
($A_{500}=-4.0 \pm 0.24$ and $A_{200}=-2.9\pm 0.18$),
which both differ by roughly $\sim 30$--$35\%$.

On average, for a given system, the iron share within $\rvir$ is slightly larger than that within $R_{500}$.
This is a consequence of the increase of $M_{\rm Fe,ICM}$ between $\rfive$ and $\rvir$, 
while the iron mass in stars remains roughly constant at $R > \rfive $ (see next Sec.~\ref{sec:fe_mass}).
More specifically, this follows the distribution of gas and stars in the clusters, with the gas mass increasing more significantly than the stellar mass going from $\rfive$ to $\rvir$, namely 
$ M_{\rm gas,200} / M_{\rm gas,500}\sim 1.8 $, 
while $ M_{*,200} / M_{*,500} \sim 1.15 $.

In principle, the conclusions drawn about the iron share calculated within given radii/overdensities are sensitive to the distributions of gas and stars, as well as to the system total mass and environment.
In the ideal case of an isolated object, the iron share at different radii should converge by enlarging the region around the centre (until the whole gas and stellar contents are taken into account).
Within a cosmic structure formation framework, however, this is difficult to be achieved due to the presence of other structures in the surrounding environment.
We note that our findings in Fig.~\ref{fig:Fe-share} are consistent with expectations about the so-called ``closure'' radius for cluster-size systems. By definition, this radius corresponds to the distance at which the baryon fraction reaches the cosmic value, thereby enclosing all the baryons of the system.
Simulation studies on cluster-size haloes show that this condition is almost fulfilled around the virial boundary or at radii corresponding to large fractions of it (despite variations in the details of different numerical modellings --- see e.g.~\citealt{ayromlou2023,angelinelli2023,rasia2025}).
To first approximation, we therefore expect that most stars that enriched the ICM and most of the enriched gas should be found within the cluster $\rvir$ by the present time, where the iron share is evaluated. This is especially the case for the high-mass end of our sample, whereas for smaller systems with $\mfive\sim 10^{14}\msun$ the outer radius examined, $\rvir$, can still be insufficient to capture all the baryons.

From an observational point of view, the iron mass in stars, $ M_{\rm Fe,*}$ is difficult to measure directly. Thus it is often estimated by means of the detected total stellar mass, $ M_*$,  assuming an average solar iron abundance, $ Z^m_{\rm Fe, \odot}$ (by mass), that is: 
\begin{equation}
    M_{\rm Fe,*} (<R) \simeq  Z^m_{\rm Fe, \odot} ~ M_* (<R). 
    \label{eq:iron_mass_obs}
\end{equation}
This means that the iron share is evaluated observationally as \cite[][]{andreon2010,ghizzardi2021}: 
\begin{equation}
    \Upsilon_{\rm Fe}^{\rm obs} (<R) = \frac{M_{\rm Fe,ICM}(<R)} { Z^m_{\rm Fe,\odot} M_{*} (<R)}.
    \label{eq:iron_share_obs}
\end{equation}
In Eq.~\eqref{eq:iron_share_obs}, $Z^m_{\rm Fe,\odot}$ is the solar iron abundance in mass fraction, namely $Z^m_{\rm Fe,\odot} = A_{\rm Fe} Z_{\rm Fe,\odot} X$. Given the solar reference by~\cite{asplund2009}, 
the Fe atomic weight $A_{\rm Fe}=55.85$, 
and a hydrogen mass fraction of $X=0.7$,
this yields $Z^m_{\rm Fe,\odot} \sim 0.00124$~\cite[following][]{ghizzardi2021,renziniandreon2014,maoz2017}.
As a general warning, we note that 
any uncertainty on the iron solar abundance adopted would thus directly translate into the same uncertainty on the iron share.
In particular,
according to Eq.~\eqref{eq:iron_mass_obs},
keeping the assumption that the stars have on average a solar iron abundance while adopting a different solar reference value would yield different stellar iron masses.

\begin{figure}
\centering
\includegraphics[width=.49\textwidth,trim=20 10 10 10,clip]{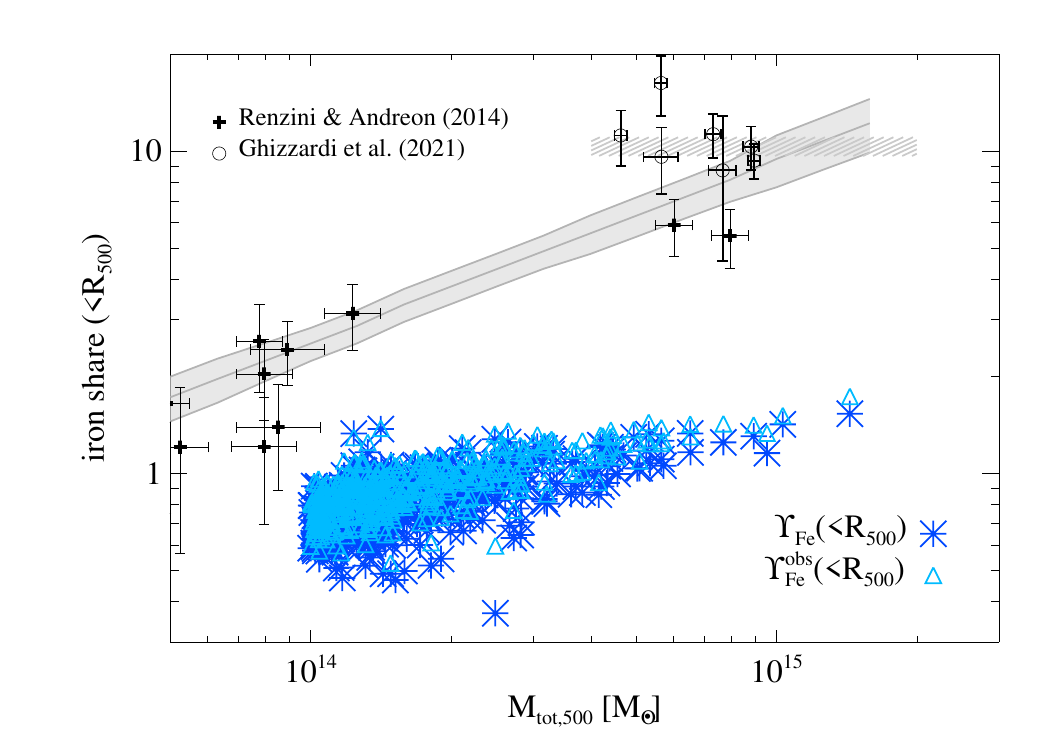}
    \caption{Comparison of the iron share from simulations and observational data by~\protect\cite{renziniandreon2014} (black crosses and solid line with grey shaded area) and \protect\cite{ghizzardi2021} (black open circles and dashed shaded area).
    The two simulation estimates of the iron share are derived by either computing directly the iron mass within stars (as in Fig.~\ref{fig:Fe-share}; blue big asterisks; "sim") or by converting the stellar mass into iron mass assuming solar abundance, similarly to observations (eq.~\eqref{eq:iron_share_obs}; cyan small triangles; "obs-like").
    \label{fig:Fe-share-obs}}
\end{figure}

In Fig.~\ref{fig:Fe-share-obs} we compare the iron share of simulated clusters within $\rfive$ against observational estimates taken from the study by~\cite{renziniandreon2014} and
from the X-COP cluster sample by~\cite{ghizzardi2021}, as a function of $\mfive$. 
To this scope, in addition to the values of $\Upsilon_{\rm Fe} (<\rfive)$ evaluated directly from the total iron mass in gas and stars and already reported in Fig.~\ref{fig:Fe-share}, we also compute the iron share via the observational approach based on Eq.~\eqref{eq:iron_share_obs}.
For the assumptions made in Eq.~\eqref{eq:iron_share_obs} on the average stellar metallicity and solar reference abundance,
we find that the observational estimate of the iron stellar mass for our simulated clusters (Eq.~\eqref{eq:iron_mass_obs}) is slightly smaller than the true value and thus the iron share slightly increases, as clear from the figure. The differences are anyhow minor, of less than $\sim 10\%$ for the majority of the systems ($75\%$ of the sample). Within $\rvir$, the median difference between true and observational-like stellar iron mass is even smaller, of the order of $\sim 5\%$
(see Appendix~\ref{app:fe_star_mass} and Fig.~\ref{fig:ironstellarmass}).

The iron share predicted from simulations is significantly smaller than what is obtained from cluster observations, at fixed total mass. 
From group to massive clusters we find a difference by a factor of $5$--$8$ --- as in Fig.~\ref{fig:Fe-share-obs}.

The value of $\Upsilon_{\rm Fe}$ in simulations confirms that gas and stars overall share a similar amount of iron, within a factor of two, despite various physical processes could impact the distributions of gas and stars in groups and clusters and their evolution in time.
For instance, if the ICM within $\rvir$ comprised an accreted gas component enriched by stars that are not any longer (e.g. splash back galaxies) or not yet within the cluster volume, then the value of the iron share $\Upsilon_{\rm Fe} (<\rvir)$ reported in Fig.~\ref{fig:Fe-share} would be overestimated compared to the expectations of the closed-box model.
Such a situation could happen when merging galaxies get stripped of their gas atmosphere, for instance after a first passage through the cluster centre, and then get shot outside $\rvir$. In terms of mass, this should be a subdominant effect, though.
Instead, if part of the gas enriched by the cluster stellar component were displaced beyond the cluster virial radius and not yet re-accreted by the present time, then the ICM iron mass would be lower than the amount produced by the stars within the cluster. As a consequence, the iron share value would be also lower, compared to the closed-box model expectation.
In this respect, a mass trend of the iron share can be partially connected to the impact of ejective feedback at early times ($z\gtrsim 2$). 
If strong, it can push enriched gas to larger distances and the depletion of iron-rich gas content is expected to be more prominent in low-mass systems, due to their shallower potential wells, than in massive clusters, where instead a larger amount of gas can be more efficiently re-accreted by the present time~\cite[e.g.][]{mitchell2022}.

Additionally, we note that part of the iron content can be comprised in a different, namely colder and denser, gas phase within the cluster, thus not contributing to the X-ray diffuse emission. 
The amount of this component is sensitive to the star-formation model in the simulations and to the thresholds adopted for the X-ray gas selection.
In our study, we find that the total iron mass within hot and cold gas components is typically $\sim 10\%$ larger (median value on the sample) than the iron mass within the X-ray gas, within both $\rfive$ or $\rvir$. The iron not included in the X-ray-emitting phase mostly resides in the core (i.e.\ $<0.2\times\rfive$) of our simulated clusters, where it typically comprises $\sim 40\%$ of the total iron gas mass and dominates the iron gas budget for $\sim 25\%$ of our clusters.

\subsection{Iron masses and spatial distribution}\label{sec:fe_mass}

The picture on the iron share outlined above finds its roots in the spatial distribution of the iron content of ICM and stars.

\begin{figure}
    \centering
    \includegraphics[width=0.99\columnwidth,trim=30 10 15 10,clip]{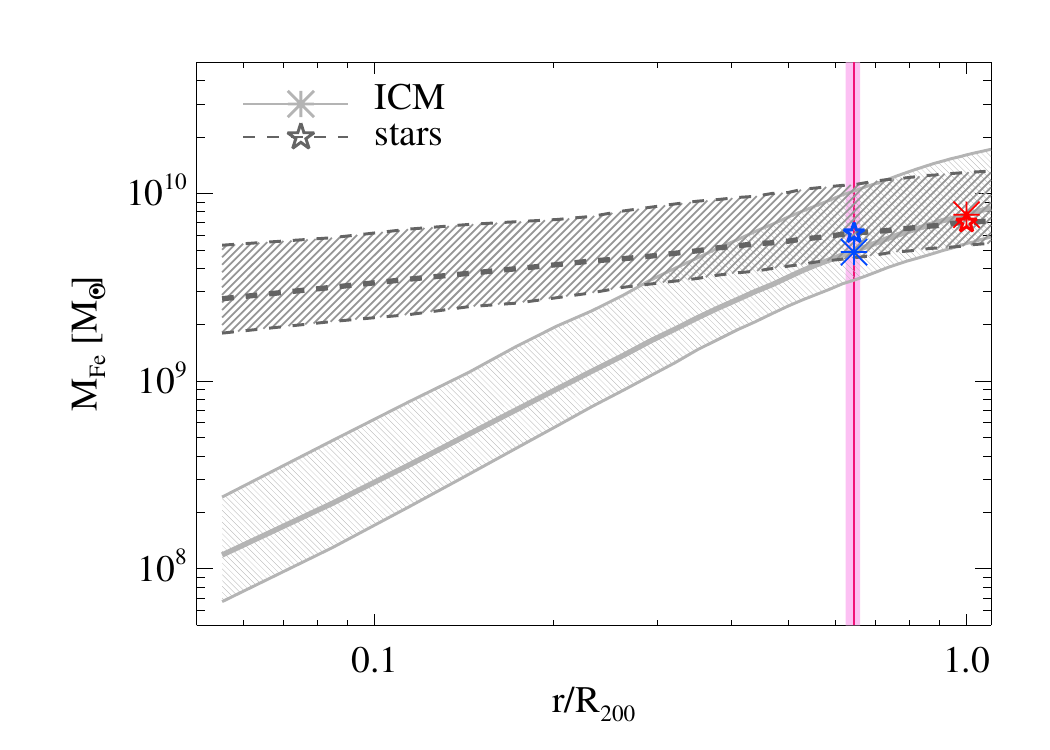}
    \caption{Median radial profiles of iron mass in the ICM (solid line) and in the stellar component (dashed line) up to $\rvir$. The scatter on the sample (data comprised between the 16$^{\rm th}$ and $84^{\rm th}$ percentile) is marked in both cases by the grey shaded area. 
    The coloured symbols mark the median value of $M_{\rm Fe}$ in the gas (asterisks) and in the stars (stars) at $\rfive$ (blue) and $\rvir$ (red) respectively.
    The vertical pink line and shaded area indicate the median value of $ \rfive / \rvir $ and the corresponding scatter.
    }
    \label{fig:fe_mass_prof}
\end{figure}

Fig.~\ref{fig:fe_mass_prof} illustrates the median radial profile (and scatter) of the cumulative iron mass in gas and stars respectively, up to $\rvir$.
From the comparison of the two median profiles, it is evident that most of the stellar $M_{\rm Fe}$ is enclosed within the innermost region ($< 0.1\times\rvir$), and roughly doubles by $\rvir$.
The cumulative $M_{\rm Fe}$ profile for the gas is instead much steeper, increasing by $\sim 1.5$ dex between $0.1\times\rvir$ and $\rvir$.
We recall that a fraction of the iron mass is not comprised in the X-ray-emitting gas considered in Fig.~\ref{fig:fe_mass_prof} but rather associated to the star-forming colder gas phase. This is however mostly located in the innermost region 
(namely $<0.2\times\rfive \sim 0.15\rvir$) and does not impact the radial trend beyond $\rfive$.
Going from $\rfive$ to $\rvir$, the increase in iron mass $M_{\rm Fe}$ is instead on average larger for the gas ($\sim 55\%$) than for the stellar component ($\sim 12\%$), similarly to the trends of
total gas and stellar masses (see also Sec.~\ref{sec:iron_share}).
We find that the relative ratio between iron mass in the ICM and total system mass is typically almost the same within both $\rfive$ and $\rvir$. Differently, for the stellar component, this ratio decreases at $\rvir$ because the total mass increases while most of the stellar iron mass (and stellar mass) is concentrated well within $\rfive$.

Despite this very different spatial distribution, the gas and stellar components share almost the same amount of iron mass within $\rfive$ or $\rvir$.
In Fig.~\ref{fig:fe_mass_prof} this is marked by the similar median values of $M_{\rm Fe}$ for gas and stars at $\rfive$ and $\rvir$ (normalised to $\rvir$ in the figure --- blue and red symbols, respectively), and is consistent with values of the iron share on average close to unity within both overdensities (see Fig.~\ref{fig:Fe-share}).

\begin{figure}
    \centering
\includegraphics[width=0.99\columnwidth,trim=30 0 20 0,clip]{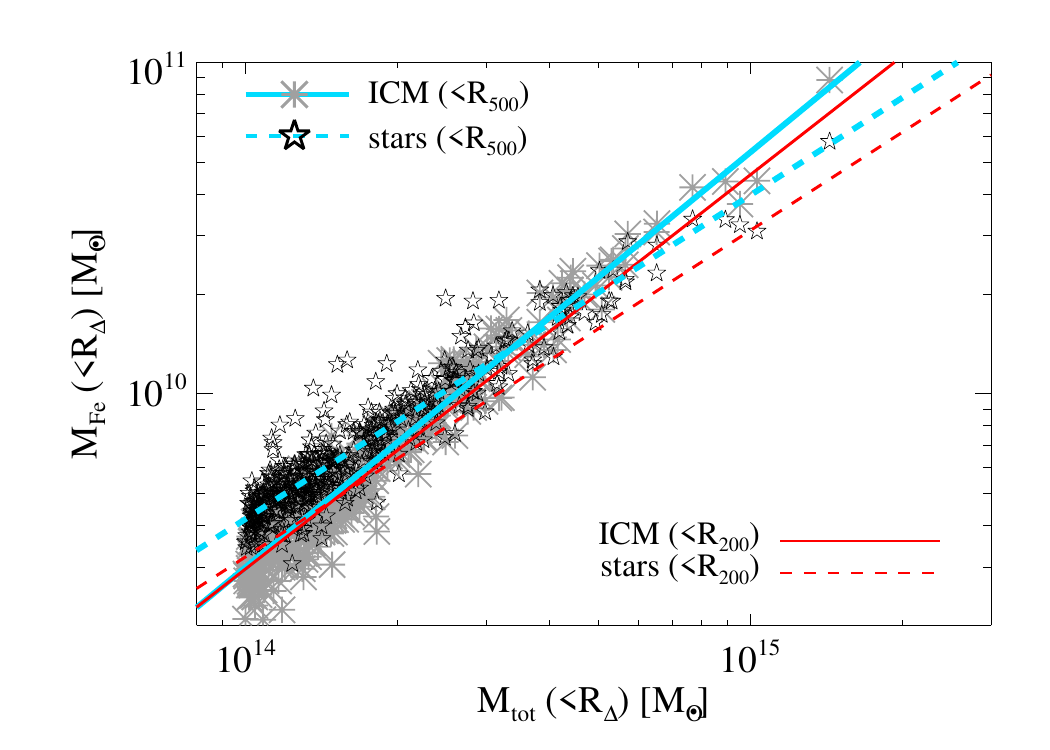}
    \caption{Relation between iron mass in gas (grey asterisks) and stars (black stars) and total mass, within $\rfive$. Lines refer to the best-fit relations between $M_{\rm Fe}(<\rfive)$--$M_{\rm tot,500}$ (cyan lines), for gas (solid) and stars (dashed) respectively.
    Overplotted for comparison, the best-fit relations for the $\rvir$ region, $M_{\rm Fe}(<\rvir)$--$M_{\rm tot,200}$ (gas: red solid line; stars: red dashed line).
    \label{fig:iron_mass_scaling}}
\end{figure}

The total iron mass in ICM and stars within both overdensities is reported as a function of the system total mass in Fig.~\ref{fig:iron_mass_scaling}.
The data points refer to the quantities computed within $\rfive$. 
We compute the log-log best-fit linear relations, in the form
\begin{equation}\label{eq:mfe_mtot_fit}
    {\rm Log}(M_{\rm Fe} / M_\odot ) = y_0 + \alpha \times {\rm Log}(M_{\rm tot} / M_\odot),
\end{equation}
for gas and stars, within $\rfive$ and $\rvir$.
Best-fit values for normalisation $y_0$ and slope $ \alpha $ are reported in Table~\ref{tab:mfe_mtot_fit}.
Both gas and stellar iron masses correlate strongly with the system total mass, and the scaling shows a relatively similar normalisation,
within both $\rfive$ and $\rvir$. In terms of slope, the relation between stellar $M_{\rm Fe}$ and total mass is always mildly shallower than the relation for the ICM iron mass.

The differences in slopes, although mild, reflect the different share of the iron content in terms of ICM and stars. This is at the origin of the $\Upsilon_{\rm Fe}$ mass trend observed in Fig.~\ref{fig:Fe-share}.
In massive systems, the gas typically comprises larger amounts of iron compared to stars ($\Upsilon_{\rm Fe} \gtrsim 1$), at both overdensities, with a larger difference at $\rvir$.
In low-mass haloes, instead, the iron content of stars is larger than that of the gas within $\rfive$ [$\Upsilon_{\rm Fe}(<\rfive) \lesssim 1$], whereas it is similar within $\rvir$ [$\Upsilon_{\rm Fe}(<\rvir) \sim 1$]. 

\subsubsection{The iron content of the ICM}
\label{sec:iron_budget_ICM}

In order to dissect the discrepancy between simulated and observed iron share, we compare the iron budget in the ICM to observational results.

\begin{figure*}
\centering
    \includegraphics[width=0.8\textwidth,trim=10 0 10 10,clip]{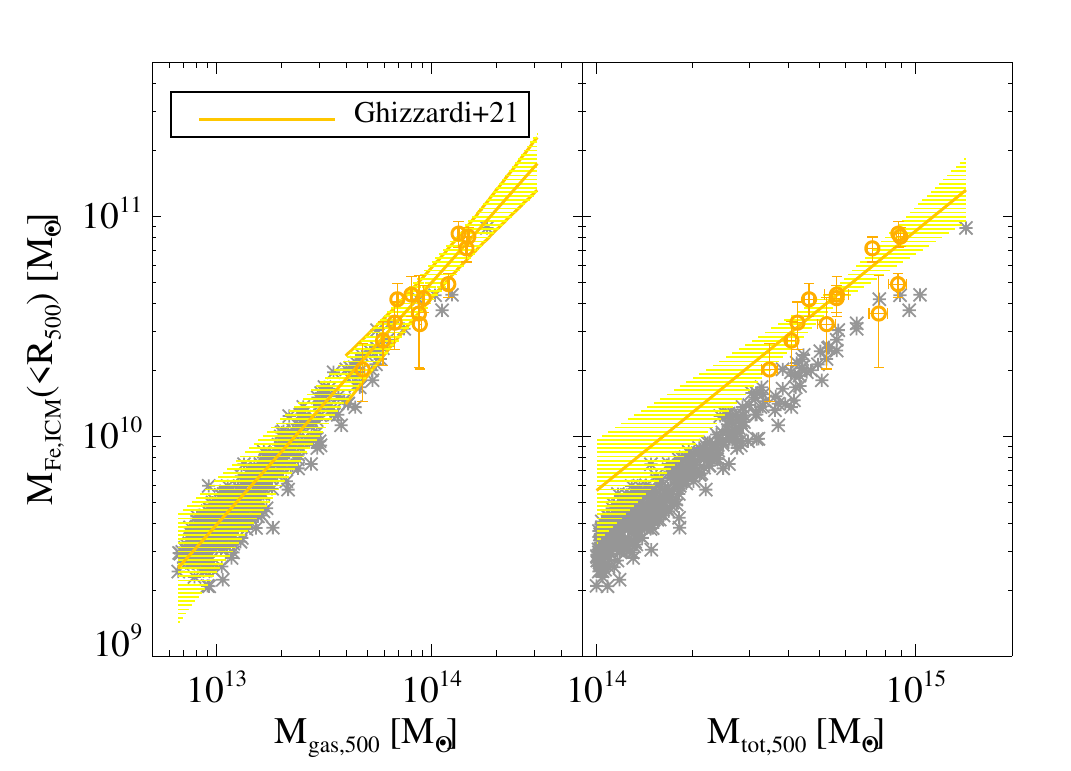}
    \caption{Iron mass in the ICM as a function of gas mass (left) and total mass (right) within $\rfive$ (grey asterisks). Orange open circles with errorbars show the results for the X-COP clusters analysed by \protect\cite{ghizzardi2021}. Best-fit observational relations from \protect\cite{ghizzardi2021} are also reported (solid lines) with $1$-$\sigma$ confidence regions (yellow shaded areas).
    \label{fig:iron_mass_gas_tot_obs}
    }
\end{figure*}

In Fig.~\ref{fig:iron_mass_gas_tot_obs} we report the relations between the ICM iron mass within $\rfive$ and both the gas mass (left) and the total mass (right) of the simulated clusters. As a comparison, we overplot the recent observational results by~\cite{ghizzardi2021} from the X-COP cluster sample, whose iron share is shown in  Fig.~\ref{fig:Fe-share-obs}.
We find consistent trends in terms of slope, for both relations.
For the total gas mass, $M_{\rm gas,500}$, we consider both the cold and hot gaseous components, while for $M_{\rm Fe,ICM}(<\rfive)$ we refer to the iron mass comprised only in the X-ray emitting gas.
In both cases we expect very similar results, though. Indeed, for the massive systems we select ($\mfive > 10^{14} \msun$), the gas budget is dominated by the hot X-ray component, which  constitutes $\gtrsim  95$\% of the total gas mass within both $\rfive$ and $\rvir$.
The relation between $M_{\rm Fe,ICM}$ and $M_{\rm gas}$ indicates that the predictions from the simulations agree remarkably well with the observational results. Although the X-COP data mainly populate the high-mass envelope of the simulated relation, the observational best-fit relation is in very good agreement with the simulated data when extrapolated to the lower-mass regime ($M_{\rm gas} < 5\times 10^{13}\msun$) as well.

In the simulated $M_{\rm Fe,ICM}$--$M_{\rm tot}$ relation we note instead a systematic offset in the normalisation, compared to data. In both axes, the offset is of order of $\sim50\%$ on average.
The X-COP iron gas masses are nonetheless still partially compatible with the simulation ones, at fixed total mass of the cluster, given the uncertainties on the measurements.
Considering the shift in the $x$-axis, we note that
observational X-COP total masses refer to hydrostatic estimates, but the estimated mass bias for this specific sample is expected to be relatively small, of order of 6\% at $\rfive$~\cite[][]{ettori2019, eckert2019}, and cannot explain the offset between the simulated and observed trends shown in the figure. 
From the simulation point of view, in contrast, we should consider a 10--20\% average shift towards lower masses if we were to estimate total masses through an observational-like approach under the assumption of hydrostatic equilibrium~\cite[see, e.g.,][for mock eROSITA observations of a sample of clusters extracted from the same Magneticum box]{scheck2023}. 
The hydrostatic mass bias thus cannot explain the observed offset of Fig.~\ref{fig:iron_mass_gas_tot_obs}~(right), nor it would impact the comparison with the iron share from observational data (Fig.~\ref{fig:Fe-share-obs}) given that the discrepancy is by almost 1~dex.
In general, we connect the shift in the normalisation of the $M_{\rm Fe,ICM}$--$M_{\rm tot}$ relation to the offset in the gas-to-total mass relation between the \Magneticum{} clusters and the X-COP sample by~\cite{ghizzardi2021} (see Fig.~\ref{fig:fgas-mtot}). 
Both simulations and observational datasets are nonetheless consistent, within the uncertainties, with the relationship proposed by~\cite{eckert2021}, which aims at defining the $f_{\rm gas}$--$M_{\rm tot}$ region occupied by the majority of recent observational findings presented in the literature.

From Fig.~\ref{fig:iron_mass_gas_tot_obs}, we conclude that the iron mass in the gas component of the simulated clusters is consistent with the one estimated for the X-COP clusters, at given total gas mass. 
This is in line with the fact that the mass-weighted iron abundance profiles for the \Magneticum{} clusters analysed here are largely consistent with those of the X-COP sample by~\cite{ghizzardi2021} (see Appendix~\ref{app:profs}).

Further ICM chemical abundances in the \Magneticum{} simulation have been previously explored by~\cite{dolag2017} across various mass ranges. They also find consistent results in comparison to a variety of observations.
These different analyses all point towards an overall consistent picture of gas chemical properties in the simulated clusters investigated in this work.
We thus explore the stellar-component properties in order to interpret our results on the iron share.

\subsubsection{The iron content of the stellar component}
\label{sec:iron_budget_stars}

For what concerns the stellar budget, we showed in Sec.~\ref{sec:iron_share} that an observational-like estimate of the iron mass in the stellar component of simulated clusters, according to Eq.~\eqref{eq:iron_share_obs}, does not impact significantly our conclusions on the (simulated) iron share (see Fig.~\ref{fig:Fe-share-obs}), since it only leads to a moderate ($\lesssim 10\%$) underestimation of the true value in most cases.
We thus inspect directly the stellar component in simulated clusters in order to single out the possible origin of the iron share discrepancy with respect to observational results.

\begin{figure}[h]
    \centering
    \includegraphics[width=0.99\columnwidth,trim=20 0 10 0,clip]{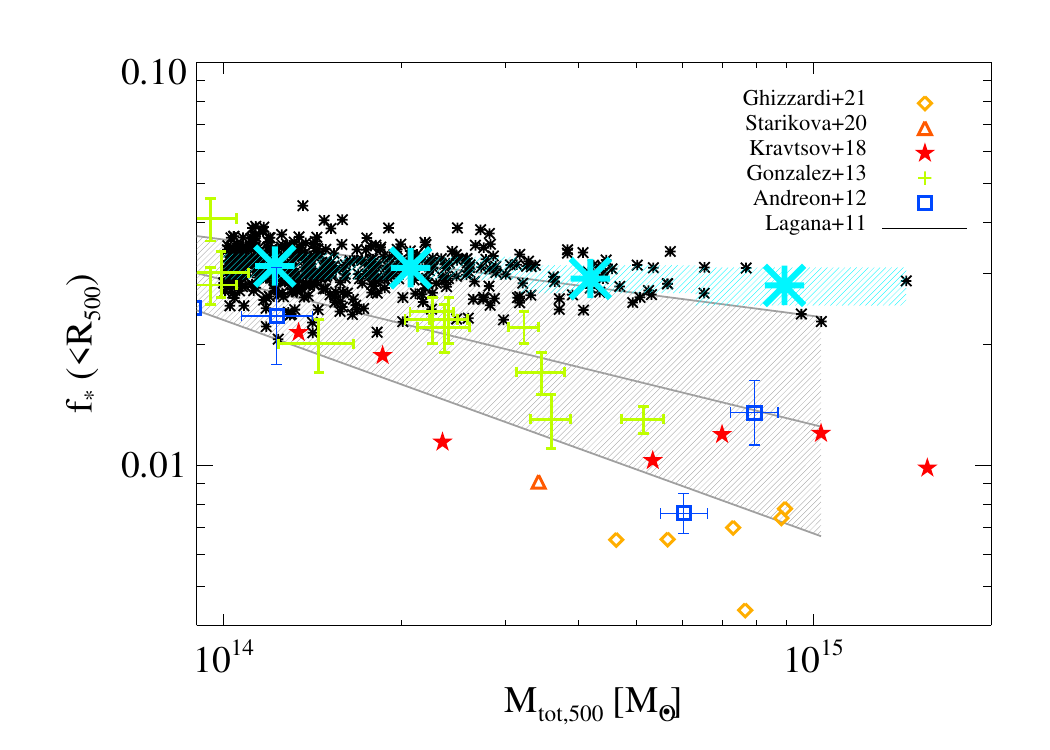}
    \caption{Relation between stellar fraction and total mass, within $\rfive$.
    Black asterisks refer to the simulated clusters in the \Magneticum{} sample, with median values and scatter in four bins of total mass marked in cyan.
    Observational data are taken from
    \protect\cite{lagana2011},
    \protect\cite{andreon2012},
    \protect\cite{gonzalez2013},
    \protect\cite{kravtsov2018}, \protect\cite{starikova2020} and~\protect\cite{ghizzardi2021}, marked as in the upper-right legend.
    \label{fig:fstar_mtot}}
\end{figure}

In the \Magneticum{} sample we find a relatively flat, mildly decreasing, relation between the stellar mass fraction and the system total mass, for both $\rfive$ and $\rvir$. For simplicity, we only show our results for $\rfive$ in Fig.~\ref{fig:fstar_mtot}.
On average, we find typical values of $f_*$ around $3\%$ within $\rfive$ and closer to $2\%$ within $\rvir$, as marked by the median values in four bins of total mass reported in the figure (cyan asterisks and shaded areas).
The intrinsic scatter of the stellar fraction is small, of the order of $0.23\%$ and $0.16\%$, at $\rfive$ and $\rvir$ respectively, 
i.e.\ $\sigma_{f_*}/f_* \lesssim 0.1$ in both cases. 
This is significantly smaller than the scatter typically measured in observed stellar fractions~\cite[e.g.][]{andreon2012, molendi2024}.

In Fig.~\ref{fig:fstar_mtot}, we compare the simulation results to observational measurements taken from \cite{lagana2011}, \cite{andreon2012}, \cite{gonzalez2013}, \cite{kravtsov2018}, \cite{starikova2020} and \cite{ghizzardi2021}. 
Overall, we note a broad agreement with the reported observational findings for the low-mass systems ($\mfive \lesssim 2$--$3 \times 10^{14}\msun$).
At the scales of massive systems ($\mfive \gtrsim 3$--$4 \times 10^{14}\msun$), simulated stellar fractions are a factor of $\sim 2$--$3$ higher than observed ones, with the majority of the data roughly converging around $f_{*}\sim 1$--$2\%$.
Among the observational datasets reported, \cite{ghizzardi2021} find the lowest values ($f_* \sim 0.7\%$ on average) for the X-COP massive clusters used to estimate the iron share discussed in Sec.~\ref{sec:iron_share}. In this case, there is a factor of $\sim 4$--$5$ difference with respect to the stellar fractions derived from our simulated clusters.
Given the stellar iron mass estimate from Eq.~\eqref{eq:iron_mass_obs}, such difference between simulations and observations in stellar fraction
would translate directly into an equal difference 
in the stellar iron mass (see Appendix~\ref{app:fe_star_mass}), and thus in the iron share.

An overall similar picture is obtained also for the measurements of the stellar component within $\rvir$ 
(as shown in Appendix~\ref{app:fstar_ficm}),
given that no strong variation in the stellar-to-total mass ratio is typically observed in cluster outskirts, $\rfive < r < \rvir$~\cite[][]{andreon2015}. Data by \cite{andreon2010, andreon2012} suggest stellar fractions decreasing from about 4\% at $ M_{\rm tot} \sim 10^{14} \,\rm M_\odot$ to 0.6\% at $ M_{\rm tot} \sim 10^{15} \,\rm M_\odot $. This is in contrast with the flatter relation around $2$--$3\%$ predicted by the simulations (see Fig.~\ref{fig:fstar_mtot200}).
A recent determination by \cite{sartoris2020} for the massive Abell~S1063 cluster ($M_{\rm tot}\sim 2.9\times 10^{15}\msun$ within a radius of $\rvir \sim 2.6$\,Mpc) at redshift $z\simeq 0.35$ reported nonetheless a stellar fraction within $\rvir$ of $ f_* = 1.5\% \pm 0.4$\%. This is higher than the figures typically inferred in the high-mass regime and is closer to simulation findings, as visible from Fig.~\ref{fig:fstar_mtot200} in the Appendix.
The value by \cite{sartoris2020} derives from a robust, full, dynamical reconstruction of the mass profile. The authors evaluate the BCG stellar mass from spectral-energy-distribution (SED) fitting (assuming a Salpeter IMF) of high-quality VLT MUSE spectra, and further account for 1234 spectroscopically confirmed cluster members up to $ \rvir $, for which stellar masses are derived from SED fitting of WFI photometric data from observations with the ESO 2.2m telescope.

We stress that results about stellar fractions from numerical simulations could be strongly impacted by the modelling of the different physical processes included (such as gas cooling and SN/AGN feedback), as well as by numerical resolution. 
Part of the difference can also be related to the treatment of the diffuse stellar component and its contribution to the total stellar mass.
In observations, the stellar content of groups and clusters of galaxies has been extensively studied, especially at low redshift ($z\lesssim 0.1$, for the majority of the works reported here). By comparing the different observational datasets, a large scatter in terms of both slope and normalisation is found~\cite[e.g., see][]{eckert2019}.
In addition to the sensitivity of the inferred stellar mass on the assumed IMF, the major source of uncertainty in observations is in fact the achievement of a comprehensive census of the stellar population out to large radii, including also the tenuous ICL component and the unresolved low-mass galaxies. 
Whereas the ICL contribution is always included in our measurements of total stellar masses from simulations (unless otherwise stated), not all the observational studies reported account for this diffuse stellar component.
For this reason, we inspect in the following the impact of measuring stellar masses within fixed apertures around the BCG instead of considering the whole stellar component within the cluster main halo.

\subsubsection{Main-halo contribution to the stellar mass}
\label{sec:mstar_estimate}

\begin{figure}
    \centering
    \includegraphics[width=0.99\columnwidth,trim=20 0 0 0,clip]{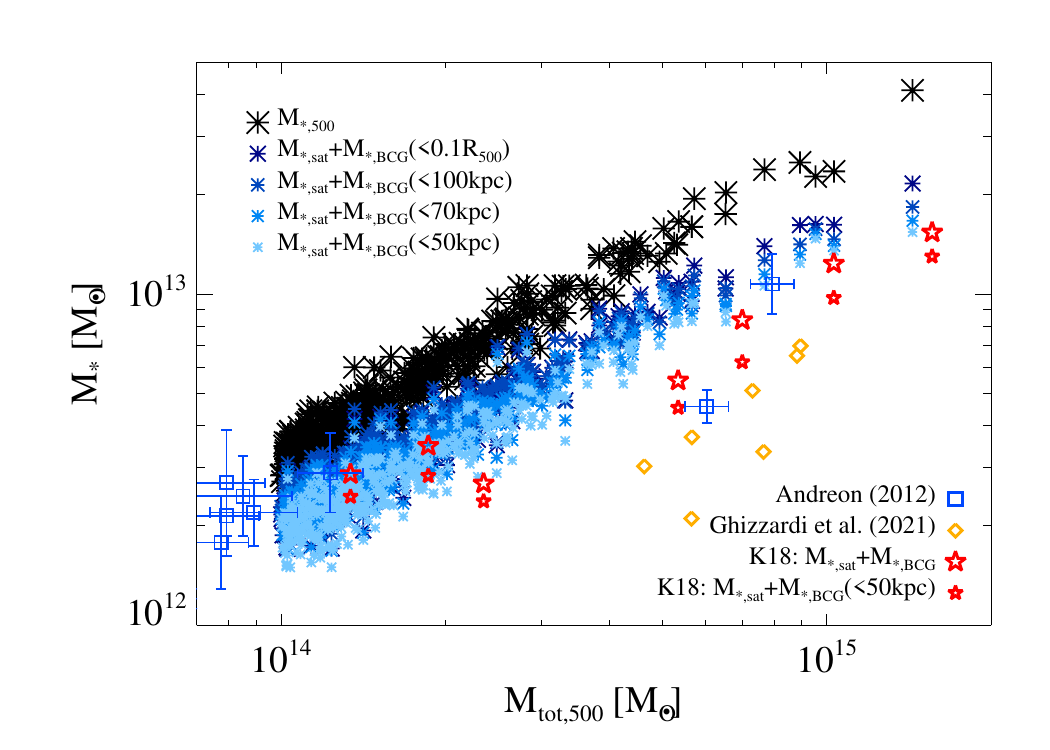}
    \caption{Scaling relations between stellar mass and total halo mass $ \mfive$, within $R_{\rm 500c}$, for simulated clusters compared to observational data.
    Simulated datapoints refer to stellar masses within $ \rfive$ ($ M_{*,500}$; black asterisks), and to values obtained by summing up the satellite stellar mass, $M_{*,\rm sat}$, and the BCG stellar mass, $M_{*,\rm BCG}$, computed within
    $0.1\,\rfive$ 
    (dark-blue asterisks),
    $100\,\kpc$
    (blue asterisks),
    $70\,\kpc$ 
    (light-blue asterisks),
    and $50\,\kpc$
    (cyan asterisks), as in the upper-left legend.
    Observational data are from 
    \protect\cite{ghizzardi2021} (yellow diamonds) and
    \protect\cite{andreon2012} (blue squares), as from their samples also the iron share estimates were derived, and from \protect\cite{kravtsov2018} (K18; red stars). For the latter, we report two estimates, given by the sum of the satellite stellar mass, $M_{\rm sat}$ (see their Tables~1 and~4), and the BCG stellar mass within $ 50\,\kpc $, $M_{\rm BCG} (<50\,\kpc)$ (small symbols), and extrapolated to infinity from their best-fit S\'ersic profile, $M_{\rm BCG}$ (big symbols).
    \label{fig:mstar_scaling}}
\end{figure}

In Fig.~\ref{fig:mstar_scaling} we show the scaling of the simulated stellar mass $ M_*(<\rfive)$ as a function of the total cluster mass $ M_{\rm tot,500}$.
We also report some of the observational datasets from Fig.~\ref{fig:fstar_mtot}, as in the legend.
In particular, we include the two datasets by~\cite{andreon2012} and~\cite{ghizzardi2021}, for which we discussed the iron share estimate in Sec.~\ref{sec:iron_share}, and the data by~\cite{kravtsov2018}.
Reflecting the results on the stellar fraction, we note a systematic overprediction of the total stellar mass from simulations within $\rfive$, for haloes with $M_{\rm tot,500} > 2 \times 10^{14}\,\rm M_{\odot}$.

Here, we investigate in detail the origin of this difference, by evaluating, in the simulated clusters, the specific contribution of BCGs and satellite galaxies, as identified by \Subfind~\cite[see also][]{ragagnin2022}.
We define as satellites the substructures of the main halo that have a stellar mass $ M_{*,\rm sat} > 10^9 M_{\odot}$ and are located within the cluster $\rfive$.
The BCG is defined as the stellar component of the main halo. With this definition, we would count all the diffuse stellar halo, extending out to large radii and making up the ICL, in the BCG stellar mass. Since this component is not included in most of the observational determination of the stellar budget, we only count the stellar mass contained within given apertures. 
In particular, we test different selection criteria for the BCG, computing the stellar mass within the following physical apertures: 
$0.1 \,\rfive $,
$100\,\kpc $, 
$70\,\kpc $, 
and 
$ 50\,\kpc $.\\
For each of these four BCG definitions, the total stellar mass of the  cluster region within $\rfive$ is computed as the sum of the stellar masses of the satellites\footnote{We verified that the difference between computing the satellite masses within fixed apertures and considering the total masses by \Subfind{} is minor.} and the BCG.
This yields four different estimates of $M_* (<\rfive)$ (coloured, smaller asterisks in Fig.~\ref{fig:mstar_scaling}) in addition to the usual one computed by summing up the mass of all the stellar particles within $\rfive$ (black asterisks).

In general, we note that the decrease in the region used to estimate the BCG mass is reflected by a decrease of the resulting $M_{*}$ which is not constant with halo mass.
The effect is stronger at larger total masses and the scaling relation effectively flattens for smaller BCG selection regions.
This can be appreciated already by considering the BCG contribution within $ 0.1 \,\rfive $.
More quantitatively, we can compare the simulation results of these tests to analogous measurements available for the observational sample by~\cite{kravtsov2018} (cf. their Table~4).
For instance, we report the data we obtain by summing up their values for satellites and BCG stellar masses within $50$ physical $\kpc$, $ M_{*,\rm sat} + M_{*,\rm BCG}(<50\,{\rm kpc})$. 
In this case, the disagreement is alleviated and the difference between simulations and observational data ranges between $10\%$ (at the low-mass end) and $40\%$ (in massive systems), at fixed total mass. 
This extreme case introduces a decrease of stellar mass by a factor of $\sim1.5$--$3$, which would directly translate into the same difference in the iron stellar mass, following Eq.~\eqref{eq:iron_mass_obs}. This would let the simulated iron share value correspondingly increase by the same factor $\sim1.5$--$3$, thus getting closer to the observational estimates (see Appendix~\ref{app:ironshare_dsc}).
This further confirms that the diffuse stellar component beyond the BCG, not comprised within satellites (either small galaxies with $ M_* < 10^{9}\,\rm M_\odot$ or diffuse stellar component of the main halo), has a significant role in shaping the level of tension between observational findings and results from our simulated sample, in which it contributes up to $60\%$ of the total stellar mass enclosed within $\rfive$.
Carefully accounting for residual systematics in the observational inference of the stellar mass budget in clusters and for the contribution of an elusive ICM component is essential to assess the limits that the feedback processes included in simulations have in regulating star formation in such extreme environments.

\subsection{Connecting star formation and ICM chemical enrichment}
\label{sec:SF_Z}

The common underlying assumption that the stars in the cluster potential well must be responsible for ICM chemical enrichment is challenged by the large values of the iron share inferred from observations of massive systems. 

A standard approach to investigate the amount of metal mass that could have been produced by the actual stellar population in the cluster galaxies and ejected in the cluster gas relies on semi-analytical methods.
In these methods, typical prescriptions about IMF, star formation efficiency and supernova ratios are usually adopted.
Studies in the literature conclude that the ICM metallicity expected from 
such calculations on the basis of the observed stellar mass should be lower than
the metallicity actually measured from X-ray spectra~\citep{bregman2010,loewestein2013,renziniandreon2014,blackwell2022}.
Here, we test semi-analytical findings by confronting the actual ICM metallicity in the simulated clusters and the predictions from stellar evolution models based on the stellar component.

In particular, we focus on the study by \cite{loewestein2013} (L13) that 
gives a relation between ICM iron abundance and stellar-to-gas mass ratio expressed as 
\begin{equation}
    Z_{\rm Fe}/Z_{\rm Fe,\odot} = \beta \times (10\,f_{*}/f_{\rm gas}),
\label{eq:L13}
\end{equation}
where the factor $\beta$ is directly proportional to the total specific number of SNe per unit stellar mass formed available to enrich the ICM, $\eta^{\rm SN}$, increases with the fraction of stellar mass returned to the environment, $r_*$, and varies with the SNIa-to-SNcc ratio, $R^{\rm SN}$ (see Sec.~2.4 of L13 for more details).
L13 derives the value $\beta=0.155$ for $\eta^{\rm SN} = 0.0046~M_\odot^{-1}$, $r_* = 0.35$, $R^{\rm SN} = 0.44$, a `diet Salpeter' IMF \cite[which is a slight modification of the Salpeter IMF that assumes a flat distribution of stellar masses below 0.6~$M_\odot$;][]{Bell2001} 
and metal yields from \cite{Kobayashi2006}.

\begin{figure*}
    \centering
    \includegraphics[width=0.99\columnwidth]{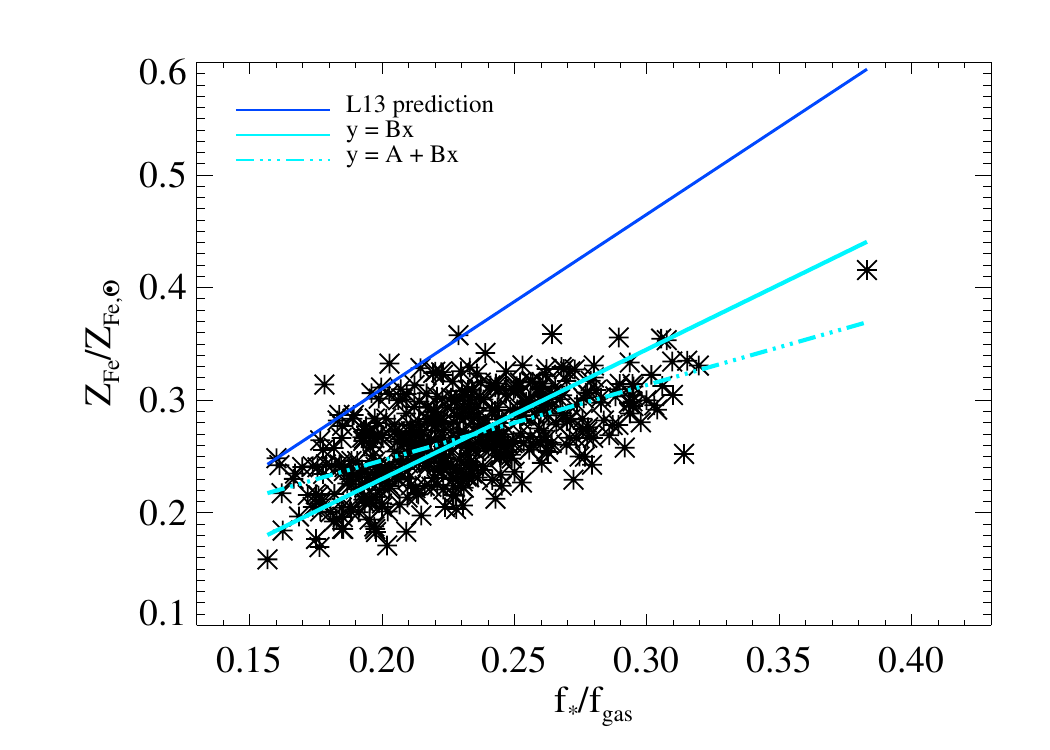}
    \centering
    \includegraphics[width=0.98\columnwidth]{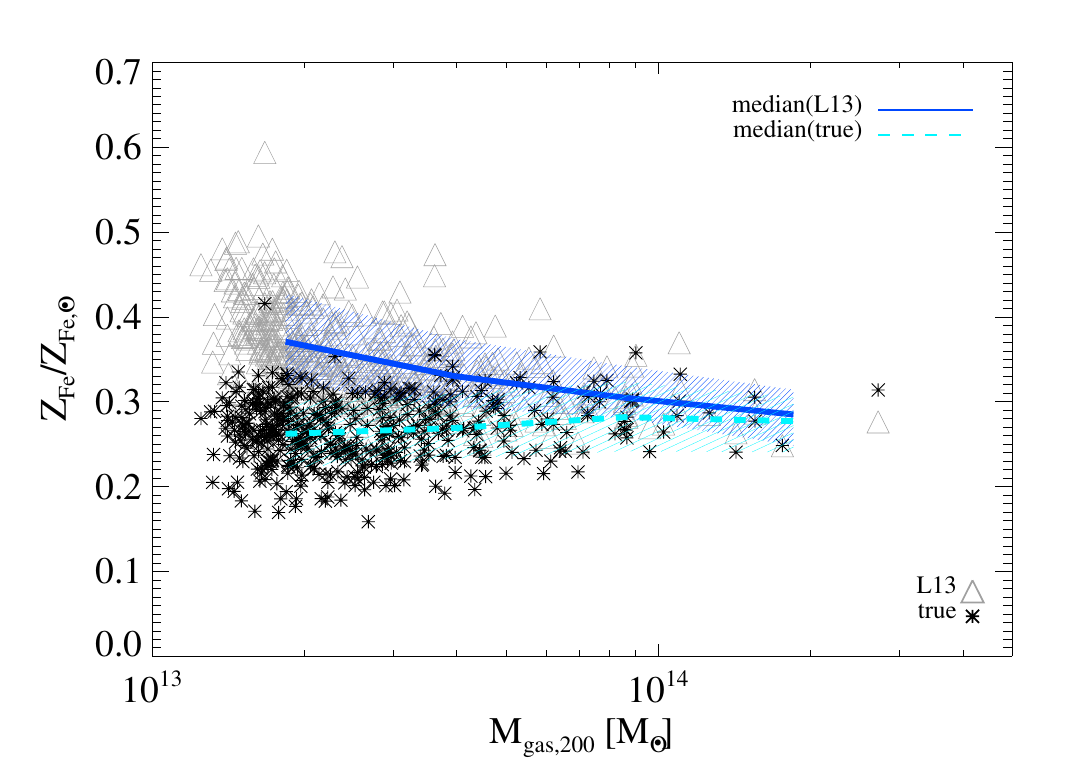} 
    \caption{Comparison between the simulated ICM mass-weighted iron abundance $Z_{\rm Fe} / Z_{\rm Fe,\odot}$ and the L13 prediction based on the stellar-to-gas fraction $f_{*}/f_{\rm ICM}$ as in Eq.~\protect\eqref{eq:L13}, for the region within $\rvir$.
    {\it Left:} relation between $Z_{\rm Fe} / Z_{\rm Fe,\odot}$ and $f_{*}/f_{\rm ICM}$ for the simulated clusters (black asterisks), compared to the L13 relation (blue solid line). Overplotted two best-fit relations (cyan solid and dot-dashed lines, as in the legend) to the simulation data.
    {\it Right:} iron abundance as a function of simulated gas masses.  
    Two estimates of $Z_{\rm Fe} / Z_{\rm Fe,\odot}$ are compared: the mass-weighted value computed directly from the simulation (``true''; black asterisks) and the value predicted for each simulated cluster given its stellar-to-gas fraction (``L13''; grey triangles).
    We mark median values in five mass bins for both the true (dashed cyan line) and the L13 (solid blue line) estimate, as well as the respective scatter (shaded areas) between the 16th and the 84th percentile of the distributions.
    \label{fig:zfe-L13}}
\end{figure*}

To better relate the discrepancy in the iron share to the stellar and gas components, we 
bracket the validity of the L13 prediction and investigate the relation between 
the ICM iron abundance and the stellar-to-gas mass ratio directly in our simulated clusters.
We compute $Z_{\rm Fe}$ both directly from the simulation outputs (``true'' value) and from the L13 prediction in Eq.~\eqref{eq:L13} (``L13'' value).
In Fig.~\ref{fig:zfe-L13}\;(left) the true mass-weighted $Z_{\rm Fe}/Z_{\rm Fe,\odot}$ is reported as a function of the 
simulated $f_{*}/f_{\rm gas}$ (black asterisks), for the region within $\rvir$.
The true ICM iron abundance mildly increases with the stellar-to-gas fraction, but the relation based on the simulation outputs is overall flatter than the one predicted by L13 in Eq.~\eqref{eq:L13} (solid blue line in the Figure).
In general, the L13 relation predicts higher ICM enrichment levels mostly at large $f_{*}/f_{\rm gas}$, mainly corresponding to small-mass systems (see discussion below and Appendix~\ref{app:fstar_ficm}).
For a more quantitative comparison, we also report the best fit to the simulation data assuming two functional forms. 
First, we adopt a linear function similar to the L13 one, i.e.\
$
    Z_{\rm Fe}/Z_{\rm Fe,\odot} = {\rm B} \times (f_{*}/f_{\rm gas}),
$
for which the best-fit value is ${\rm B} = 1.15$ (corresponding to $ \beta = 0.115 $).
Then, we test a second functional form: 
\begin{equation}
    Z_{\rm Fe}/Z_{\rm Fe,\odot} = {\rm A} + {\rm B} \times (f_{*}/f_{\rm gas}),
\label{eq:fit2}
\end{equation}
for which we find the best fit for $A=0.11$ and $B=0.67$.
 Thanks to the additional offset parameter $A$, equation~\eqref{eq:fit2} provides a better fit (smaller residuals) to simulation results.
These fits are valid in the range $ 0.15 \lesssim f_{*} / f_{\rm gas} \lesssim 0.40 $.

In our simulated clusters, at each simulation timestep and for each stellar particle, $\eta^{\rm SN}$, $R^{\rm SN}$ and $r_*$ derive directly from our time-dependent stellar-evolution calculations (that include both metal spreading and stellar-mass loss during AGB phases and SNIa/SNcc events; see Sect.~\ref{sec:sims}).
We do not impose nor calibrate any relation between star formation (or metallicity) and $f_{*} / f_{\rm gas}$.
Simulated stellar mass fractions result consistently from the gas thermal cooling/heating and environmental properties.
The metal distribution is additionally affected both by dilution due to mergers and by feedback mechanisms, that can halt star formation temporarily and impact the spreading of heavy elements away from star-forming regions.

Differences, albeit modest,
between the L13 relation and the one found in simulations can also be introduced by different assumptions in the stellar evolution model.
The choice of the IMF also plays a role during the metal budget buildup \cite[e.g.][]{tornatore2007}.
For instance, the number of SNe per unit stellar mass is sensitive to the chosen IMF, but stellar masses decrease only by a factor of $ \sim 1.24 $ when passing from a diet Salpeter IMF to a Chabrier IMF \citep{bell2003, gallazzi2008, herrmann2016}.
Commonly adopted values for the stellar return fraction typically range between $r_* \simeq 0.26$ and $r_* \simeq 0.45$ for different assumptions on stellar metallicity, IMF and metal yields~\cite[e.g.][]{Fardal2007, ORourke2011, Vincenzo2016}.
Here, we have verified that the $z=0.07$ median return fraction predicted from our simulated stellar populations in the clusters is $r_* \sim 0.41$, in line with the expectations for a Chabrier IMF (which is adopted in our simulation runs and which is used to evaluate SN quantities at each timestep for each stellar population) and fairly close to the diet Salpeter assumption in L13.
The chosen stellar yields affect only slightly the overall normalisation of $Z_{\rm Fe}$ and the total metallicity \cite[e.g.][]{tornatore2007, maio2010, buck2021}.
Therefore, details in the stellar-evolution modelling  have little impact on improving the almost 1-dex discrepancy between observations and simulations in the iron share.
Overall, since L13-based gas metallicities and simulation-based gas metallicities differ by only a factor of $\lesssim 1.5$ at all $f_\star / f_{\rm gas}$ ratios and since the expected ICM iron mass is roughly consistent with observational data (as shown e.g. in previous Fig.~\ref{fig:iron_mass_gas_tot_obs}), the much larger discrepancy in the iron share has to be mostly related to the different (higher) $f_{*} / f_{\rm gas}$ regime spanned by our clusters.

As a further check in our analysis, we inspect in Fig.~\ref{fig:zfe-L13}\;(right)
the discrepancy between true iron abundance (black asterisks) and L13 prediction (grey triangles) as a function of gas mass.
For both L13 expectations and simulation predictions, the scatter on the iron abundance is of about $0.02$--$0.04\,Z_{\rm Fe,\odot}$ over the whole cluster mass range.
This is essentially introduced by the scatter in the relation between $f_{*}/f_{\rm gas}$ and mass of the simulated clusters (see Appendix~\ref{app:fstar_ficm}).
Overall, the true ICM iron abundance is fairly similar for all halos in our mass range, with a shallow mass dependence that is in agreement with most simulation results already reported in the literature from group to cluster regime~\cite[e.g.][]{fabjan2010, dolag2017, truong2019, pearce2021, nelson2024}.
Compared to the true abundances, the values predicted according to L13 are on average $\sim 34\%$ higher. This difference is in fact mass dependent, with larger discrepancies for poor clusters and groups (roughly $40\%$) than for massive clusters (few percents).
This mass trend is due to the different impact of feedback in different mass regimes.
Indeed, in the shallower potential wells of small systems not all the baryons involved in the metal cycle are enclosed within $\rvir$ by the present time and part of the metal-rich gas is likely displaced out of $\rvir$.
This turns into an ICM enrichment level similar to massive systems, but lower than expected (according to the L13 prediction) from their larger $f_{*}/f_{\rm gas}$ values measured within $\rvir$ (see the decreasing, albeit shallow, trend of $f_{*}/f_{\rm gas}(<\rvir)$ with total mass $\mvir$ in Fig.~\ref{fig:fstar-ficm}).
At variance with observations, the true ICM metallicity in the simulated massive clusters ($M_{\rm gas,200} \gtrsim 5\times 10^{13}\msun$) is fairly consistent with the L13 expectation based on the stellar-to-gas mass fraction.
In the high-mass end of our sample, true and L13-predicted ICM iron abundances differ by less than $\sim 10\%$, further stressing that the simulated ICM enrichment level is essentially consistent with predictions from semi-analytical stellar evolution models.
As commented above, the small differences seem thus to be related to the different modelling techniques.
Namely, changes in the amount of metals produced can be due to small variations in the assumptions underlying the L13 estimate
compared to the \Magneticum{} simulations,
as well as to the impact of 
highly non-linear processes followed by the simulations (such as star formation, feedback mechanisms and accretion/merging events) throughout the whole cluster evolution.
All these phenomena impact the enrichment process and contribute in a non-trivial way to, for instance, the spreading of metal-enriched gas, the manner metals are reincorporated into newly-formed stars and the spatial distribution of gas and stars within systems of different mass.
The relation with the total gas mass in Fig.~\ref{fig:zfe-L13}\;(right) further indicates that all these effects 
impact more the low-mass (high-$f_\star / f_{\rm gas}$) regime than the massive clusters, where instead the iron-share discrepancy is the largest.
This means that the much bigger offset between the observationally inferred iron share and the corresponding theoretical expectations cannot be easily reconciled with simple model variations, but is rather dependent on differences in the stellar-to-gas relative contributions to the total mass budget.

\subsection{Effective iron yield}
\label{sec:iron_yield}

Lastly, we inspect the effective iron yield, which is another way of quantifying the efficiency with which stars produce iron in clusters.
It is defined as the ratio between the total iron mass contained in cluster gas and stars ($M_{\rm Fe, gas}+M_{\rm Fe, *}$) and the initial mass of the cluster stellar component ($M_{*,0}$):
\begin{equation}\label{eq:iron_yield}
    y_{\rm Fe} = \frac{M_{\rm Fe, gas}+M_{\rm Fe, *}}{M_{*,0}}.
\end{equation}
This can also be expressed in solar units as $y_{\rm Fe,\odot}=y_{\rm Fe}/Z_{\rm Fe,\odot}^m$ and evaluated within a spherical region of radius $R_\Delta$.
The initial stellar mass, $M_{*,0}$, is typically larger than the present-day mass of the stars in the cluster ($M_*$) due to stellar mass losses during the temporal evolution of the stellar populations, namely
\begin{equation}\label{eq:mstar_ini}
    M_* = (1-r_*) \times M_{*,0}\, ,
\end{equation}
for an effective stellar mass return fraction $r_*$. 
In our simulations, 
the initial mass of every stellar particle is stored in every snapshot output, so that we can directly compute the total $M_{*,0}$ for the stars located in the clusters. 
This further allows us to verify the cumulative mass loss of the stars in the clusters, where multiple stellar populations of different ages co-exist and the whole cosmic evolution has occurred by the late-time snapshot considered.
We find that the effective return fraction for our clusters by $z=0.07$ is in fact consistent with $r_* \sim 0.41$ (as expected theoretically for a Chabrier IMF, adopted in our runs).
This is also fairly close to the value adopted in observations, where $M_{*,0}$ has to be estimated from Eq.~\eqref{eq:mstar_ini} by assuming a value for $r_*$ (e.g. $M_* = 0.58 \times M_{*,0}$, i.e.\ $r_*=0.42$, both in~\citealt{renziniandreon2014} and~\citealt{ghizzardi2021}).

\begin{figure}
    \centering
    \includegraphics[width=0.99\columnwidth,trim=40 10 10 20,clip]{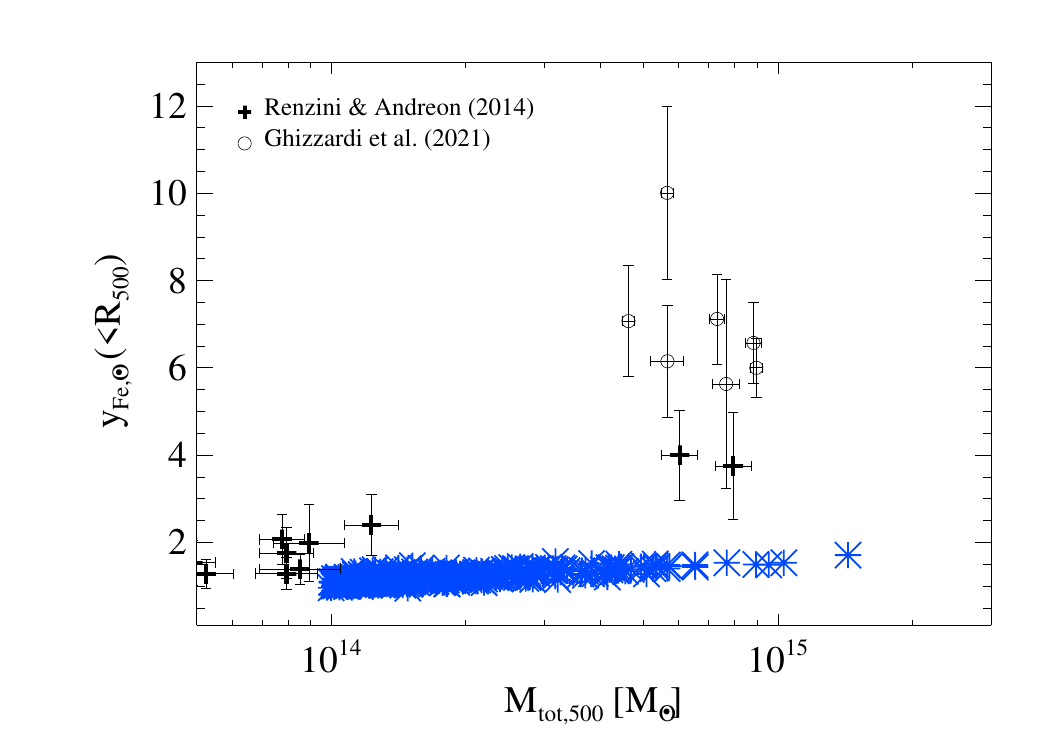}
    \caption{Effective iron yield as a function of cluster mass, within $\rfive$. Simulation data are marked by blue asterisks, and compared to observational data by~\protect\cite{renziniandreon2014} (black crosses) and \protect\cite{ghizzardi2021} (black open circles).}
    \label{fig:iron_yield}
\end{figure}

Simulation results on the effective iron yield are displayed in Fig.~\ref{fig:iron_yield} as a function of total mass, within $\rfive$. For comparison, we also report observational estimates by~\cite{renziniandreon2014} and~\cite{ghizzardi2021}, as in Fig.~\ref{fig:Fe-share}.
Consistently with the picture sketched by the iron share, in the simulations we find a flatter relation between $y_{\rm Fe}$ and system mass.
This translates into a difference by a factor of 
$\sim 3$--$5$ between simulated and observed effective iron yield (depending on the observational dataset) in massive systems ($\mfive \gtrsim 4 \times 10^{14}\msun$), with simulations indicating a lower efficiency in the iron production by the cluster stars. 

From Eqs.~\eqref{eq:iron_yield} and~\eqref{eq:mstar_ini}, we note that the quantities involved in the effective iron yield are essentially the same at play in the iron share, so that they make a consistent description of the iron production and circulation between stars and gas. 
The discrepancy between simulated results and observational estimates is thus ascribed to the same factors, mostly to the different stellar mass to total mass relation (on which both stellar iron mass and initial stellar mass depend) and, to a minor extent, to some difference in the gas-to-total mass relation (see Secs.~\ref{sec:iron_budget_ICM} and~\ref{sec:iron_budget_stars}, and the following discussion in Sec.~\ref{sec:discussion}).
In particular, these differences consist in the overprediction of both the total iron mass in massive clusters ($M_{\rm Fe, gas}+M_{\rm Fe, *}$) and the initial stellar mass ($M_{*,0} = M_* /(1-r_*)$) in simulations. 
According to Eq.~\eqref{eq:iron_yield}, the size of the discrepancy in the iron yield between simulations and observations is therefore partially alleviated compared to the iron share case.

In the following, we thus concentrate our discussion on the iron share result, which in principle is a quantity more directly measurable in observations as well, by considering only the present-day gas and stellar content.

\section{Discussion} \label{sec:discussion}

The excess of observed iron abundance with respect to the predictions from stellar-evolution models, based on the observed stellar content, represents the main inconsistency in the picture of the cluster enrichment suggested by observations.
In general, the relation between iron share and total mass is expected to be impacted by the way stellar and gas distributions vary as a function of the total cluster mass. 

For the galaxy clusters extracted from a \Magneticum{} simulation box, we find a flatter relation between $f_*$ and mass, compared to observations, with simulation predictions in line with observed results only at the scale of poor clusters, while predicting too much stellar mass in the most massive systems. 
The flat trend of $f_{*}$ with mass in our sample is consistent
with findings on the stellar-to-halo mass relation in the regime of clusters ($M_{\rm tot} > 10^{14}\msun$) based on halo occupation distribution models
\cite[][]{shuntov2022}. 
Furthermore, we find a mildly increasing trend of $f_{\rm gas}$ with total mass, as shown in Fig.~\ref{fig:fgas-mtot}, which makes the dependence of the stellar-to-gas fraction, $f_*/f_{\rm gas}$, on total cluster mass shallow, especially within $\rvir$  (see Appendix~\ref{app:fstar_ficm}). 
These results and the shallow dependence of $Z_{\rm Fe,gas}$ on $f_*/f_{\rm gas}$ also reflect the almost flat trend of $Z_{\rm Fe,gas}$ with total mass found in simulated clusters~(\citealt{fabjan2010,dolag2017, biffi2018Rv, truong2019, pearce2021}; see also recent observational results by~\citealt{yates2017, mernier2018Rv, gastaldello2021, mernier2022}). 
These considerations explain why the iron share of simulated clusters is only mildly increasing with the cluster total mass, and robustly predict a much flatter trend than reported by observational investigations. 
Differently, the precise value of the iron share strongly depends on the exact level of enrichment and on the total gas and stellar budget.
In our study, the discrepancy in the value of simulated and observed iron share in massive clusters can be explained by the combination of two main factors. Each of them inevitably depends on the specific simulation and observational dataset used for the comparisons.
Considering the gas budget, iron abundances and iron masses are broadly consistent between the \Magneticum{} clusters and several observational samples, at given total gas mass.
With respect to the X-COP dataset, used for comparisons of the iron share, the scaling between gas iron mass and gas total mass in simulated clusters agrees very well in terms of both normalisation and slope. The agreement is in fact found throughout the mass range explored, not only at the high-mass end where the two datasets overlap.
There is however an offset in the gas-to-total mass relation (see Fig.~\ref{fig:iron_mass_gas_tot_obs} and Appendix~\ref{app:fstar_ficm}) which actually induces an additional factor of $\sim 1.5$ difference in the gas iron mass, and thus in the iron share, at fixed total cluster mass. 
Given such difference, the bias likely affecting the observational hydrostatic mass estimates would not be sufficient to reconcile the results. In particular, an average mass bias of 10--20\% applied to the \Magneticum{} masses to mimick the observational measurements would not significantly ameliorate the results on the iron share nor on $M_{\rm Fe,ICM}-M_{\rm tot}$ relation, and we can thus safely ignore it.
The dominant contribution to the iron share discrepancy between simulations and X-COP data is due to the difference by a factor of $\sim 5$ in stellar mass, and hence in the resulting stellar iron mass, at fixed halo mass.  
\cite{ghizzardi2021} draw a somewhat similar conclusion based on the systematics affecting their observational measures. Namely, they argue that the uncertainties affecting the ICM iron mass estimate are the smallest ones, while systematics on the stellar component are dominant and impact the resulting iron share the most.

In light of these results, reducing the discrepancy between simulations and observations in the stellar-to-total mass relation would largely alleviate the mismatch on the iron budget. 
While differences in the normalisation can always be due to uncertainties in the underlying stellar evolution assumptions, a flatter $f_{*}$--$M_{\rm tot}$ relation, as expected from theoretical studies, would definitely help reconciling observations with the predicted iron yield in the cluster regime~\cite[and iron share;][]{renziniandreon2014,molendi2024}.
From the simulation point of view, further work on the star formation modelling is also required.
Currently, most simulation results tend to overpredict the stellar masses in cluster-size haloes, producing too massive BCGs and overly massive main halos (namely the sum of the BCG and the diffuse stellar component), compared to observations~\cite[see e.g. recent results by][]{teklu2017, bahe2017, pillepich2018, henden2020, bassini2020, ragagnin2022, nelson2024}.
Compared to observations, the stellar mass function for the \Magneticum{} simulations at redshift $z=0$ predicts too many galaxies both at the low-mass end and at the largest masses, whereas a good agreement is found in the intermediate regime~\cite[at masses $\sim 10^{10}$--$10^{11}\msun$;][]{dolag2025}.
Also the star formation history in BCGs from simulations adopting similar physical modelling as the \Magneticum{} ones is still partially in contrast with observational findings, suggesting a residual star formation activity at late times which is not detected in observations of low-$z$ BCGs~\cite[][]{ragone2018}.
On the other hand, a suppression of star formation in simulations should be achieved by maintaining an efficient production and circulation of metals, in such a way to preserve the agreement between the observed and the simulated ICM enrichment level.

An additional source of discrepancy between simulated and observational measurements of stellar masses can be the comparison method itself~\cite[][]{munshi2013,brough2024}. In fact, in both cases, a conversion between stellar light and mass must be adopted, requiring unavoidable assumptions on the cluster stellar population properties, the IMF, the star formation history~\cite[][]{behroozi2010}.
From the purely observational point of view, a large variance in the measurement of stellar masses from various datasets is intrinsically present, as visible from Fig.~\ref{fig:fstar_mtot}.
Especially in massive systems, uncertainties on galaxy stellar masses and difficulties in detecting and estimating the fainter diffuse stellar component likely lead to an underestimation of the observed masses, and consequently of the iron budget in stars.
The definition of ICL in observations is ambiguous {\it per se} and different methods provide different results compared to estimates from simulations, where the identification of the diffuse stellar component is also non-trivial~\cite[][]{dolag2010}.
\cite{remus2017} showed that the diffuse stellar component around massive galaxies within group and cluster environments in the \Magneticum{} simulations reproduces the spatial distribution of the observed ICL as well as its kinematic properties.
Testing the observational approaches on mock images of simulated clusters extracted from different simulations, including \Magneticum{}, \cite{brough2024} recently showed that on average all simulations provide consistent results, with BCG+ICL fractions of $0.38\pm 0.16$. Compared to the simulation values, they find all observational measures to be biased towards underestimating the ICL fractions (yielding BCG+ICL fractions of $0.13 \pm 0.05$ on average). The authors also point out that among different methods and different simulations the range of BCG+ICL fractions obtained is actually substantially large, with several aspects such as projection effects contributing to enhance the uncertainties. Nonetheless, \cite{brough2024} conclude that the discrepancy between observations and simulations mainly resides in observers and simulators often measuring different quantities, while inherent differences related to the specific simulation model are subdominant in this respect. 

The underestimation of the observed stellar mass in massive haloes is indicated as part of the origin of the iron conundrum also by recent studies. 
\cite{ghizzardi2021} discuss various possible uncertainties affecting the calculation of iron yield and iron share, concluding the the most prominent ones are associated to the contribution of the ICL to the total cluster luminosity (up to $\sim 50\%$) and to the systematic errors in the stellar mass estimates (up to $\sim 60\%$).
\cite{molendi2024} propose an observation-based model for the chemical enrichment of clusters which implies a flat relation between the iron yield and the halo mass.
Despite favouring a flat relation between stellar fraction and total mass, the main difference between the model proposed by~\cite{molendi2024} and the simulation-based picture outlined here lies in a lower normalisation for massive systems, implying an underestimation of the stellar masses in observational analyses by $\sim30\%$, much lower than would be required to have consistency between observational results and predictions from our simulated clusters.

\section{Summary and Conclusions}\label{sec:conclusion}

In this paper we have directly investigated a long-standing problem in the metal budget share between ICM and stars in massive galaxy clusters by analysing a large set of 448 simulated systems with masses $\mfive > 10^{14}\msun$ extracted from the \Magneticum{} simulation set at $z=0.07$.
The cosmological hydrodynamical simulations include a sub-resolution treatment of a large variety of physical processes (see Sec.~\ref{sec:sims}), including a detailed chemical evolution model following a metallicity-dependent cooling, a sub-resolution model of star formation from a multi-phase ISM and the production and release of chemical elements from stars into the surrounding gas~\cite[][]{tornatore2004,tornatore2007}.
This allowed us to test whether the stars included within simulated clusters at the present epoch are enough to explain the ICM enrichment, finding a positive answer. 
On the one hand, this may sound an obvious outcome of the modelling within simulations. However, these results further ensure that the assembly process and evolution of galaxy clusters --- including merging events, stellar and AGN energy feedback, accretion of pristine gas along with pre-enriched gas, accretion of galaxies or galaxy groups --- 
do not significantly alter the final baryonic and metal budget in massive clusters compared to expectations from a close-box model. If the iron share is evaluated within a sufficiently large region, comparable to the virial one, its value is close to unity.
This simulation prediction is in contrast with observational findings that indicate a much larger iron share and thus an iron conundrum: the amount of iron detected in the ICM through X-ray observations exceeds significantly the amount that cluster stars could have reasonably produced given their total amount as inferred from optical/near-IR luminosities.

In the following, we briefly summarise our main results.
\begin{itemize}
    
    \item The iron share $\Upsilon_{\rm Fe}$ in simulated clusters is a shallow function of total mass, both within $\rfive$ and $\rvir$. Typical values are always comprised between 0.5 and 2 in the mass range explored, with $1<\Upsilon_{\rm Fe}<2$ for the most massive systems (see Fig.~\ref{fig:Fe-share}).
    
    \item Compared to observational estimates of the iron share by~\cite{renziniandreon2014} and~\cite{ghizzardi2021}, the iron share from simulations within $\rfive$ is $\sim 8$ times smaller for massive systems ($\mfive > 5\times 10^{14}\msun$), and the relation with total mass significantly flatter (see Fig.~\ref{fig:Fe-share-obs}).
    Calculating the iron mass in stars by assuming a constant average solar iron abundance for all stars does not improve the comparison significantly.

    \item The relations between total mass and iron mass in the ICM and in stars are quite similar, with slightly different slopes (Fig.~\ref{fig:iron_mass_scaling} and Appendix~\ref{app:mfe-mtot_fitBCES}). At fixed total mass, stars comprise more iron mass than ICM in the poor-cluster regime, whereas the opposite is true for massive clusters. This can be ascribed to a larger gas depletion in smaller systems, which yields iron share values lower than unity especially within $\rfive$.
    
    \item The iron content in ICM and stars within simulated clusters has a different spatial distribution, with the stellar iron mass being more centrally concentrated. In the ICM, the iron mass increases substantially with radius, when going from $0.1\times\rvir$ to $\rvir$, as the gas is dominating the outskirts baryonic budget. As a consequence, the increase between $\Upsilon_{\rm Fe}(<\rfive)$ and $\Upsilon_{\rm Fe}(<\rvir)$ is mainly driven by the increase of the ICM (gas and) iron mass.

    \item The ICM chemical properties and iron mass in simulated clusters are overall consistent with observations (Sec.~\ref{sec:iron_budget_ICM} and Appendix~\ref{app:profs}). 
    With respect to the X-COP clusters by~\cite{ghizzardi2021}, the $M_{\rm Fe, gas}$--$M_{\rm gas}$ relation shows a remarkably good agreement in terms of both slope and normalisation.
    The agreement is preserved also at the lower masses, if the best-fit relation derived from the X-COP clusters is extrapolated.
    The scaling between iron mass in the ICM and total cluster mass, within $\rfive$, is also consistent with observations in terms of slope but we find a shift in the normalisation that can be connected to a higher gas fraction, at fixed total mass, for the X-COP clusters compared to our simulated sample (see Appendix~\ref{app:fstar_ficm}). 
    This contributes a factor of $\sim 1.5$ to the difference between simulated and observed iron share.

    \item 
    The stellar fraction-to-total mass relation in simulations is flatter than what is indicated by observations, with a typical scatter of $\sim 0.2$\,dex at fixed total mass.
    Simulated massive clusters have larger total stellar masses (fractions) at fixed total mass, compared to observed systems.
    This difference in the stellar component is the main driver of the iron share gap between simulations and observations, contributing a factor of 2--5 discrepancy.
    
    \item The prediction of ICM metallicity based on the stellar-to-gas mass fraction, as in the semi-analytical approach by~\cite{loewestein2013}, yields values that are relatively close to the true ICM Fe abundance measured in the simulated clusters at the largest halo masses, while larger discrepancies between predicted and true values are found in lower-mass systems. This relates to a flatter relation between $Z_{\rm Fe}$ and $f_{*}/f_{\rm gas}$ in the simulated clusters. The impact of this different relation is found to be more important than the different underlying stellar evolution model assumptions.

    \item As indicated by the comparison between the effective iron yield in simulations and observations, we find a lower efficiency with which stars produce iron in simulated clusters, with values $1 \lesssim y_{\rm Fe,\odot} (<\rfive) \lesssim 2$ and a shallow mass dependency for $10^{14} < \mfive [\msun] \lesssim 2\times 10^{15}$.
    
\end{itemize}

Our results point out the most important directions for future investigations in this framework. 

From the simulation point of view, the scaling relation between stellar mass and halo mass definitely requires further improvements, given the overestimation of stellar masses in the regime of massive systems (both for individual BCG and global stellar masses), a result shared by independent sets of simulations.
In terms of star formation activity, BCGs in clusters extracted from simulations adopting a similar physical modelling as the \Magneticum{} ones
also show some residual star formation at low redshift which is in contrast with observations. 
Observational data at higher redshifts will thus be crucial to better constrain the star formation activity in BCG across cosmic time and to improve the modelling.
Any new model dedicated at improving the description of the galactic component should nonetheless preserve the solid results currently obtained on the ICM properties, such as global scaling relations, spatial distribution and chemical enrichment properties.

On the observational side, a better understanding of the systematics affecting the measurement of stellar masses and ICL contribution in the low surface brightness limit is definitely needed to reduce the uncertainties, and to better assess by how much current star formation and feedback models implemented in simulations eventually need to be revised. The new era opened by {\it Euclid} and {\it JWST} observations will be of paramount importance in this respect.

As a last note, a dedicated effort should still be put into properly comparing simulated and observational quantities for the stellar mass determination, in order to reduce as much as possible any biases related to the method of comparison itself.

\begin{acknowledgements}
The Authors would like to thank the anonymous referee for constructive comments that helped improving the manuscript.
VB is particularly grateful to Silvano Molendi, Sabrina De Grandi and Stefano Andreon, for fruitful discussions on the observational findings about the iron share conundrum.
VB would like to acknowledge the precious help of Camilla Maio in spotting a typographical error in Fig. C.1 during the revision phase.
VB, UM, ER acknowledge partial support from the INAF Grant 2023 ``Origins of the ICM metallicity in galaxy clusters". 
ER and VB acknowledge support from the Chandra Theory Program (TM4-25006X) awarded from the Chandra X-ray Center which is operated by the Smithsonian Astrophysical Observatory for and on behalf of NASA under contract NAS8-03060.
UM was supported in part by grant NSF PHY-2309135 to the Kavli Institute for Theoretical Physics (KITP).
KD acknowledges support by the COMPLEX project from the European Research Council (ERC) under the European Union’s Horizon 2020 research and innovation program grant agreement ERC-2019-AdG 882679.
This research was partially supported by the Excellence Cluster ORIGINS and by the Munich Institute for Astr-, Particle and BioPhysics (MIAPbP) which are funded by the German Research Foundation (Deutsche Forschungsgemeinschaft, DFG) under Germany's Excellence Strategy – EXC-2094 – 390783311, and by the DFG project nr. 415510302.
The calculations for the hydro-dynamical simulations were carried out at the Leibniz Supercomputer Center (LRZ) under the project pr83li (\Magneticum{}). 
This work has been initiated under the Project HPC-EUROPA3 (INFRAIA-2016-1-730897), with the support of the EC Research Innovation Action under the H2020 Programme; in particular, the author gratefully acknowledges the computer resources and technical support provided by CINECA.
SB is supported by: the Italian Research Center on High Performance Computing Big Data and Quantum Computing (ICSC), project funded by European Union - NextGenerationEU - and 
National Recovery and Resilience Plan (NRRP) - Mission 4 Component 2, within the activities of Spoke 3, Astrophysics and Cosmos Observations; the National Recovery and Resilience Plan (NRRP), Mission 4,
Component 2, Investment 1.1, Call for tender No. 1409 published on
14.9.2022 by the Italian Ministry of University and Research (MUR),
funded by the European Union – NextGenerationEU– Project Title
"Space-based cosmology with Euclid: the role of High-Performance
Computing" – CUP J53D23019100001 - Grant Assignment Decree No. 962
adopted on 30/06/2023 by the Italian Ministry of Ministry of
University and Research (MUR); the INFN Indark Grant.
\end{acknowledgements}

\bibliographystyle{aa}
\bibliography{mybib}

%-------------------------------------------------------------------
\begin{appendix}

\section{Iron abundance profiles}\label{app:profs}

\begin{figure}
    \centering
    \includegraphics[width=0.99\linewidth,trim=30 15 10 10,clip]{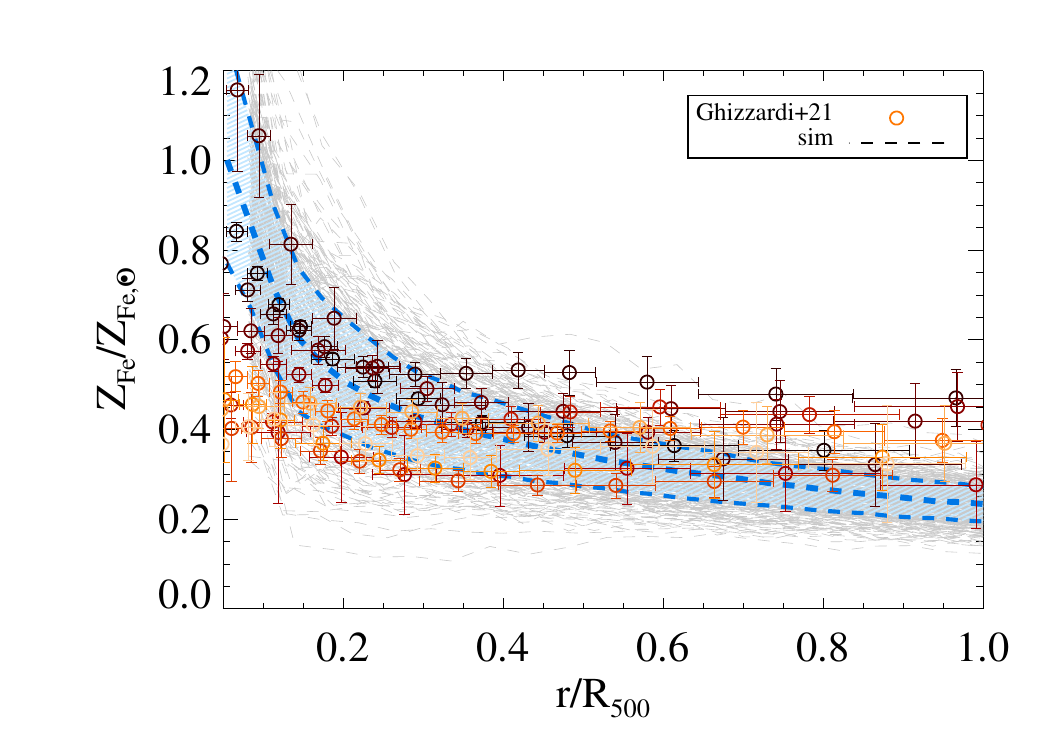}
    \caption{Iron abundance profiles up to $\rfive$ for the individual \Magneticum{} sample (thin dashed grey lines). Simulated median profile and scatter are marked as blue thick dashed line and shaded area. Coloured datapoints refer to the observational profiles for individual X-COP clusters by~\protect\cite{ghizzardi2021}, with normalisation rescaled to the reference solar abundance by~\cite{asplund2009}, used for the profiles of the simulated clusters. 
    }
    \label{fig:fe_profs}
\end{figure}

In Fig.~\ref{fig:fe_profs} we show the iron abundance profiles for the \Magneticum{} sample compared to the X-COP observational sample by~\cite{ghizzardi2021} (dark-to-light red data-points), used for comparisons on the iron share.
For the 448 simulated clusters, we report the individual mass-weighted, three-dimensional profiles (thin grey lines), as well as the sample median profile with scatter (dark grey, dotted line and shaded areas).
Compared to the X-COP clusters, we note that on average the simulated profiles are slightly steeper and reach a lower median value in the outermost region ($r\gtrsim 0.8\rfive$).
In the core, we find moderately higher abundances in simulations, but we note that this is still in agreement with some of the systems in the X-COP sample.
Considering the scatter on both samples, the simulated mass-weighted profiles are overall consistent with the observed ones throughout most of the radial range.
This is in line with the agreement found on the relation between total iron mass in the gas and total gas mass (see Fig.~\ref{fig:iron_mass_gas_tot_obs}).

\section{Iron mass scaling relations}\label{app:mfe-mtot_fitBCES}

\begin{figure}
    \centering
    \includegraphics[width=0.99\linewidth,trim=25 10 10 10,clip]{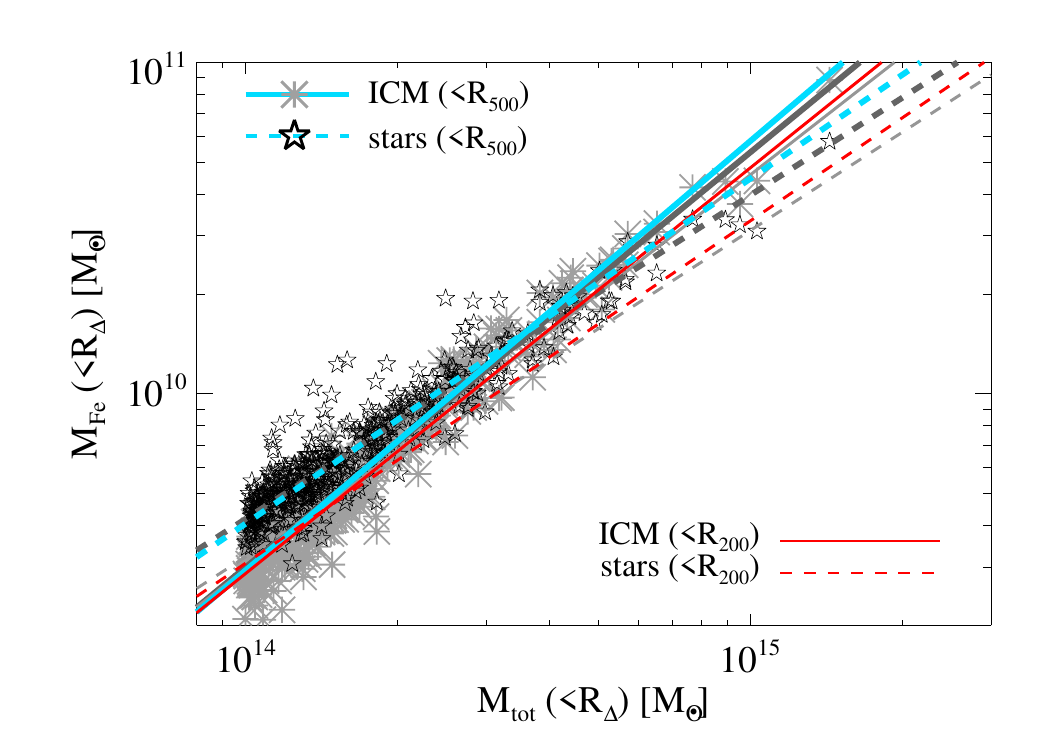}
    \caption{Same as Fig.~\protect\ref{fig:iron_mass_scaling}, where the best-fit relations have been computed using the BCES method instead. For comparison the best-fit lines from Fig.~\protect\ref{fig:iron_mass_scaling} are also reported in dark ($<\rfive$) and light ($<\rvir$) grey.}
    \label{fig:iron_mass_scaling_BCES}
\end{figure}

In Fig. \ref{fig:iron_mass_scaling_BCES} we report the results of two different fitting methods applied to the scaling relations between iron mass, $M_{\rm Fe}$, in gas (solid lines) and stars (dashed lines), and the system total mass, $M_{\rm tot}$. 
As in Fig.~\ref{fig:iron_mass_scaling}, we only show the data points for the region within $\rfive$, whereas the best-fit scaling relations are reported for both $\rfive$ and $\rvir$ overdensities (in blue and red, respectively).
Specifically, we compare the best-fit relations obtained in Sec.~\ref{sec:fe_mass} by minimising the chi-square (grey lines) with those obtained through the BCES~\cite[][]{akritas1996} linear regression method, using the least-squares bisector method as in~\cite{isobe1990}.

The best-fit slopes and normalisations, for both methods and overdensities are reported, with errors, in Table~\ref{tab:mfe_mtot_fit}.
\begin{table*}
    \centering
    \caption{\label{tab:mfe_mtot_fit}Best-fit values of normalisation $y_0$ and slope $\alpha$ of the $M_{\rm Fe}$--$M_{\rm tot}$ relation.}
    \begin{tabular}{c|c|c|c|c|c}
        & & \multicolumn{2}{c|}{$< \rfive$}  & \multicolumn{2}{c}{$< \rvir$} \\
            \hline
        & & $y_0$ & $\alpha$ & $y_0$ & $\alpha$ \\  
        \hline
        gas & linfit      & $-8.05 \pm 0.23$ & $1.25 \pm 0.02$ &  $-7.18 \pm 0.20$ & $1.19 \pm 0.01$\\
            & BCES-bisect & $-8.67 \pm 0.23$ & $1.30 \pm 0.98$ &  $-7.68 \pm 0.20$ & $1.22 \pm 0.92$\\
        \hline
        stars & linfit      & $-4.06\pm 0.24$& $0.98 \pm 0.02$ & $-4.28\pm 0.21$ & $0.98\pm 0.01$ \\
              & BCES-bisect & $-4.97\pm 0.25$& $1.04 \pm 0.79$ & $-4.97\pm 0.20$ & $1.03\pm 0.79$ \\
    \end{tabular}
    \tablefoot{The best-fit relations for gas and stars at the two considered overdensities, expressed as in Eq.~\eqref{eq:mfe_mtot_fit}, have been obtained by either minimising the chi-square statistics (``linfit'', as in Fig.~\ref{fig:iron_mass_scaling}) or via the BCES-bisector method (``BCES-bisect''), and are all displayed in Fig.~\protect\ref{fig:iron_mass_scaling_BCES}.}
\end{table*}
The two sets of best-fit relations are largely overlapping out to $(6$--$7)\times 10^{14}\msun$, with slopes consistent within the errors. 
Minor differences are mainly driven by the few most-massive systems ($M_{\rm tot} \gtrsim 8 \times 10^{14}$), and do not affect the qualitative conclusions on the main trends presented in Sec.~\ref{sec:fe_mass}.

\section{Stellar iron content}\label{app:fe_star_mass}

In the simulated clusters, we verified that the iron stellar mass estimate according to Eq.~\ref{eq:iron_mass_obs} does not differ significantly from the direct estimate obtained by summing up the iron content of all stellar particles residing within the considered region.
In particular, the observational-like estimate is on average lower than the simulation value, with a relative difference $1-M_{\rm *,Fe}^{\rm obs-like}/M_{\rm *,Fe}^{\rm true}$ that is $\lesssim 10\%$ for the majority of our clusters. This difference does not depend significantly on the total cluster mass,
featuring only a very weak anti-correlation quantified by a Person correlation coefficient of $-0.15$.
\begin{figure}
  \includegraphics[width=.98\columnwidth,trim=20 10 10 10,clip]{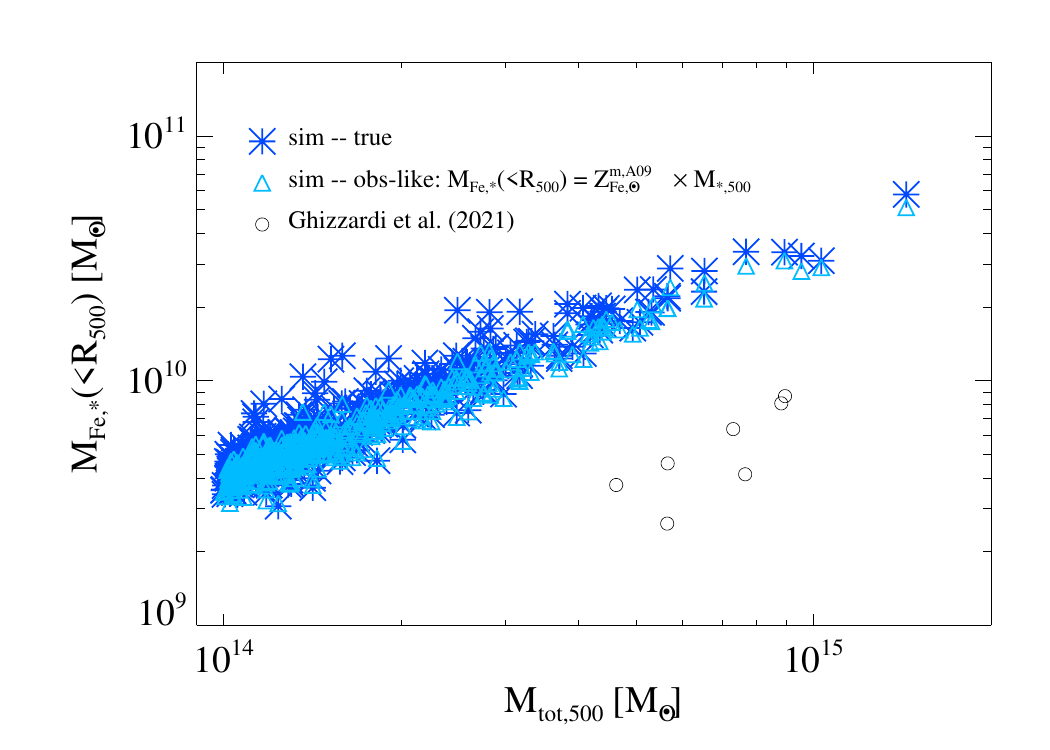}     
    \caption{Stellar iron mass as a function of the total mass for the {\sc Magneticum} sample. We report two estimates of the iron mass in the stellar component, computed by directly summing up the iron content of all stellar particles within $R_{500}$ (blue asterisks) and according to Eq.~\protect\eqref{eq:iron_mass_obs} (cyan triangles). For comparison, we report the X-COP observational data by~\protect\cite{ghizzardi2021}.
    \label{fig:ironstellarmass}}
\end{figure}
This can be seen from Fig.~\ref{fig:ironstellarmass}, where the two estimates of the stellar iron mass in the simulated clusters is shown as a function of $\mfive$.
Given this overall agreement between the two estimates, the stellar iron mass to total mass relation features the same scaling as the stellar to total mass relation (see Fig.~\ref{fig:mstar_scaling}).
Compared to the observational estimates by~\cite{ghizzardi2021} for the X-COP sample (open circles) we thus find that simulated iron masses are consistently larger in simulations.

\section{Impact of the diffuse stellar component on the iron share}\label{app:ironshare_dsc}

\begin{figure}
    \centering
    \includegraphics[width=0.96\linewidth,trim=25 10 10 10,clip]{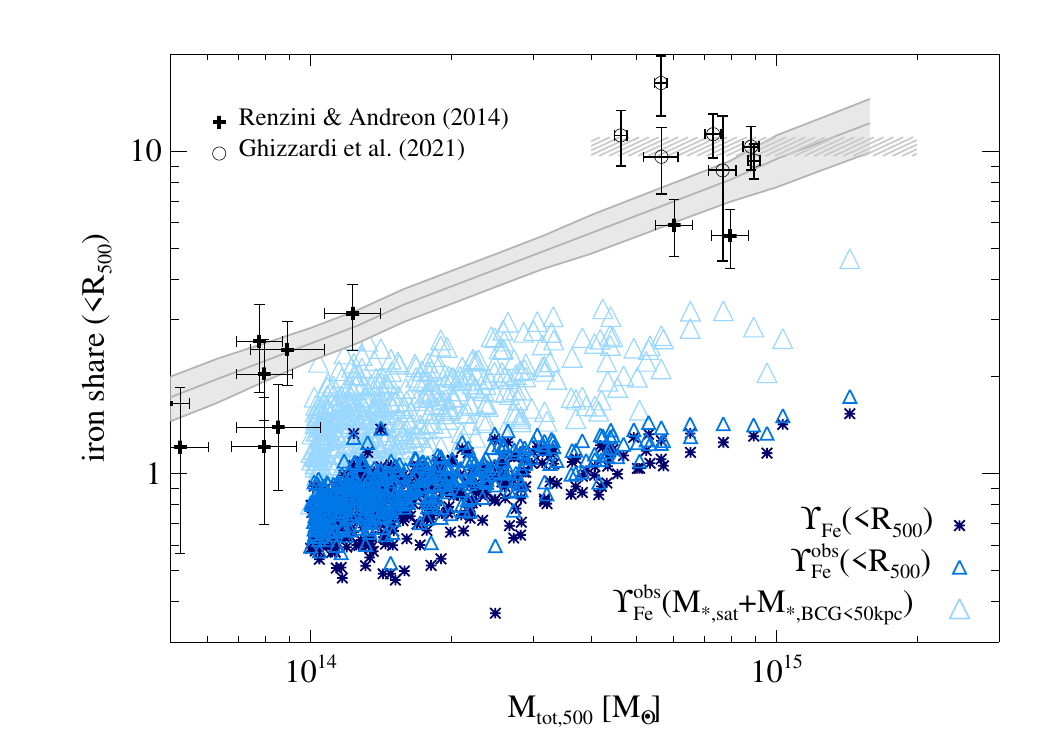}
    \caption{Same as Fig.~\ref{fig:Fe-share-obs}, but also including the additional dataset (light-blue bigger triangles) that refers to a stellar iron mass that includes only stars within the satellites and the BCG, with the latter defined within the central $50$\,kpc (smallest, light-blue asterisks in Fig.~7 of the main paper).
    \label{fig:iron_share_dsc}}
\end{figure}

Here we show the impact on the iron share of excluding the contribution of the diffuse stellar component of the main halo beyond the BCG from the total stellar mass of the cluster, and thus from the stellar iron content.
As an example, we consider the maximum correction to the stellar mass in our simulated clusters reported in Fig.~\ref{fig:mstar_scaling}.
Namely, we consider only the stars in the clusters comprised within satellites and BCG, with the latter computed within the central $50$ physical kpc ($M_{\rm *,sat} + M_{\rm *,BCG}(<50\,{\rm kpc})$, light-blue smallest asterisks in Fig.~\ref{fig:mstar_scaling}). The stellar iron mass, for this test, is then estimated as in Eq.~\eqref{eq:iron_share_obs} assuming an average solar iron abundance by~\cite{asplund2009} for all those~stars.
In Fig.~\ref{fig:iron_share_dsc} we report the iron share of simulated and observed clusters, as in Fig.~\ref{fig:Fe-share-obs}, with the additional dataset of simulation values for the aforementioned stellar mass definition (light-blue big triangles). The difference in cluster stellar mass directly translates in the stellar iron mass and thus on the iron share value. The latter increases by a factor of $\sim 2$, reducing the gap with the observed values.

\section{Stellar and gas fraction}\label{app:fstar_ficm}

\begin{figure}
    \centering
    \includegraphics[width=0.95\columnwidth,trim=20 10 10 10,clip]{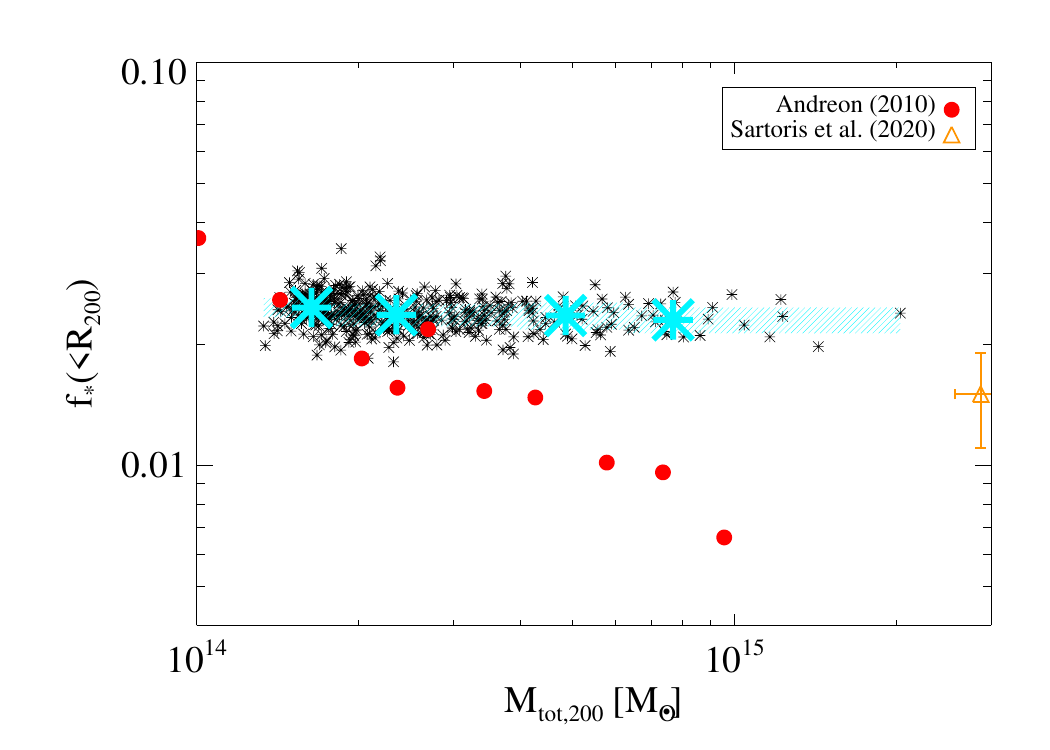}
    \caption{Relation between stellar fraction and total mass, for the region within $\rvir$. Simulation data for the \Magneticum{} sample are marked as black asterisks, with cyan big asterisks and shaded area indicating the median values and scatter in four bins of total mass. Observational data are taken from~\cite{andreon2010} (red filled circles) and~\cite{sartoris2020} (orange triangle).}
    \label{fig:fstar_mtot200}
\end{figure}

In Fig.~\ref{fig:fstar_mtot200} we report the relation between stellar fraction and total mass for the region enclosed within $\rvir$. Simulation data for the \Magneticum{} sample are compared against observational measurements by~\cite{andreon2010} and~\cite{sartoris2020}.
Simulation results show a picture that is in line with findings at $\rfive$ (cf.\ Fig.~\ref{fig:fstar_mtot}), indicating a flatter relation compared to observations, and higher values of the stellar fraction, at fixed total mass, in the regime of massive systems. We note however that the observational measurement by~\cite{sartoris2020} indicates a larger stellar-to-total mass fraction ($\sim 15\%$, for a system with $\mvir \sim 2.9\times 10^{15}\msun$), definitely closer to the simulation~\mbox{values}.

\begin{figure}
    \centering
    \includegraphics[width=0.95\columnwidth,trim=20 10 10 10,clip]{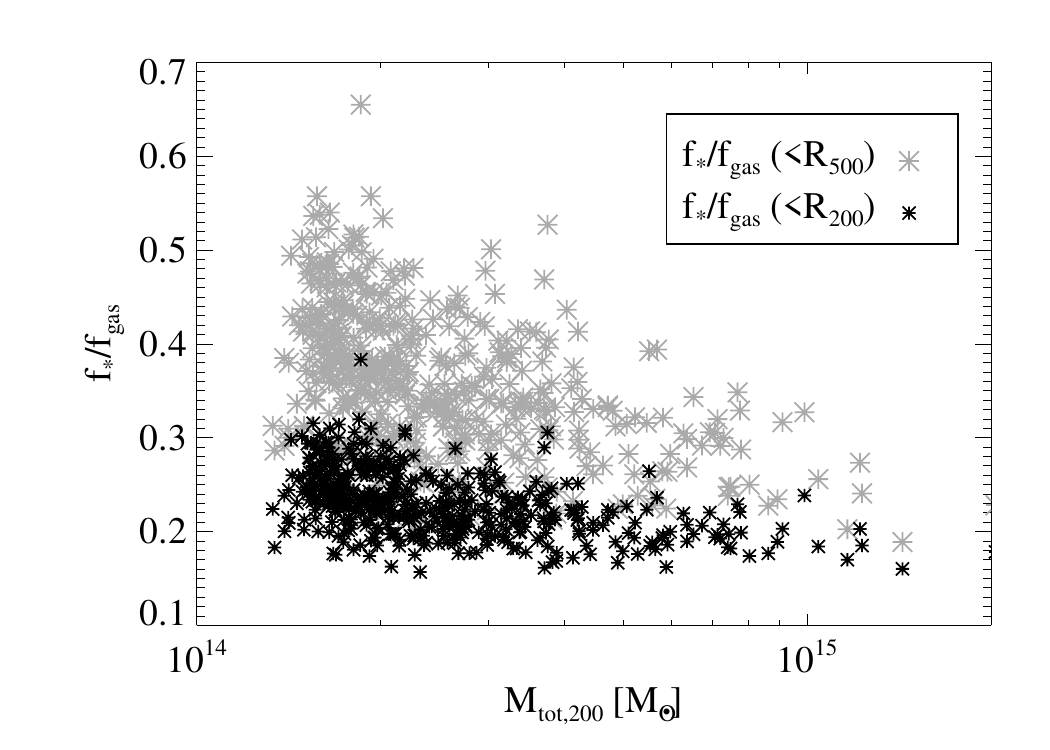}
    \caption{Stellar-to-gas fraction within the two considered overdensities ($\Delta=500$ and $\Delta=200$), both as a function of the total mass within $\rvir$.}
    \label{fig:fstar-ficm}
\end{figure}

In Fig.~\ref{fig:fstar-ficm} we report the stellar-to-gas mass fractions within $\rfive$ and $\rvir$ as a function to the system total mass $\mvir$, for the sample of \Magneticum{} clusters.
We note that there is an intrinsic scatter in both relations, which is however strongly reduced when the $f_*/f_{\rm gas}$ ratio is also computed within the larger region corresponding to $\rvir$.
The latter is in fact expected to enclose most of the system baryons, especially in massive systems, while a larger halo-to-halo variation persists at lower masses. In smaller objects, in fact, baryonic physical processes have a stronger impact in shaping the stellar and gaseous distributions.
Overall, we note that simulation values for $f_*/f_{\rm gas}$ are typically larger than observed values~\cite[cf., for instance,][]{blackwell2022}. This tension is mostly driven by the larger stellar fractions predicted by simulations in massive clusters.

\begin{figure}
    \centering
    \includegraphics[width=0.99\columnwidth]{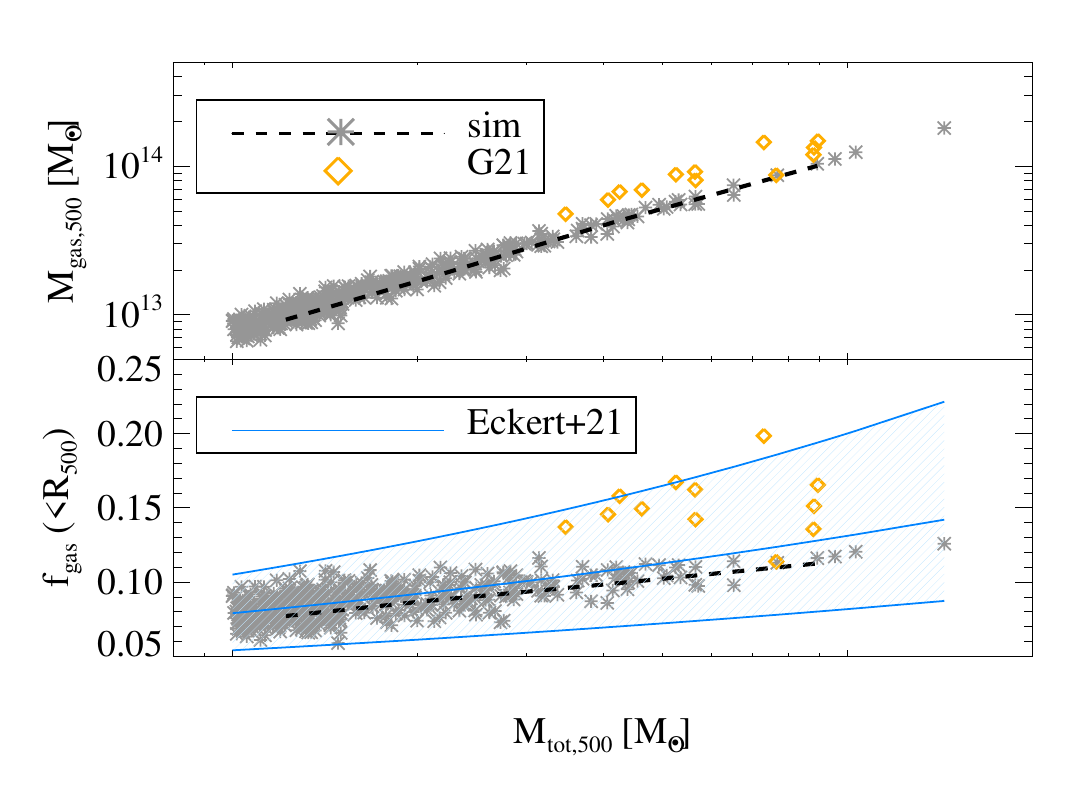}
    \caption{Upper panel: Relation between gas mass and total mass. Lower panel: Gas fraction as a function of total mass. 
    The \Magneticum{} data are reported as grey asterisks, with median values in four bins of $\mfive$ marked by the dashed black line.
    Observational data points by~\protect\cite{ghizzardi2021} (orange diamonds) and functional form derived from state-of-the-art available observations by~\cite{eckert2021} (blue line and shaded area).}
    \label{fig:fgas-mtot}
\end{figure}

We report in Fig.~\ref{fig:fgas-mtot} gas masses (upper panel) and fractions (lower panel) as a function of total mass within $\rfive$, for the simulated \Magneticum{} clusters in comparison to the X-COP sample by~\cite{ghizzardi2021}. In order to mark the scatter in observational measurements, we also report the best-fit functional form proposed by~\cite{eckert2021} (see their Eq.~11). As illustrated in Sec.~\ref{sec:iron_budget_ICM}, we note that the gas content in the X-COP sample is typically larger than in our simulated clusters, at fixed total mass.

\FloatBarrier
\end{appendix}

\end{document}